%% file: paper.tex
\documentclass[twocolumn]{aastex6}

\usepackage{xcolor}
\usepackage{amsmath}
\usepackage{affils}

\newcommand{\apx}{\ensuremath{\sim}}
\newcommand{\incomment}[1]{{\color{red} #1 }}
\newcommand{\lxlb}{\ensuremath{L_\text{X}/L_\text{bol}}}
\newcommand{\msun}{M$_\odot$}

\newcommand{\paeight}{\ensuremath{R_{80}}}
 \newcommand{\pr}{\ensuremath{R_s}}
\newcommand{\prot}{\ensuremath{P_\text{rot}}}
\newcommand{\teff}{\ensuremath{T_\text{eff}}}

\newcommand{\gps}{\ensuremath{g_\text{P1}}}
\newcommand{\rps}{\ensuremath{r_\text{P1}}}
\newcommand{\ips}{\ensuremath{i_\text{P1}}}
\newcommand{\zps}{\ensuremath{z_\text{P1}}}
\newcommand{\yps}{\ensuremath{y_\text{P1}}}
\newcommand{\gri}{(\textit{g}--\textit{r}--\textit{i})$_\text{P1}$}
\newcommand{\izy}{(\textit{i}--\textit{z}--\textit{y})$_\text{P1}$}
\newcommand{\riz}{(\textit{r}--\textit{i}--\textit{z})$_\text{P1}$}

\begin{document}
\shortauthors{Kado-Fong et al.}
\submitted{}
\title{M~Dwarf Activity in the Pan-STARRS 1 Medium-Deep Survey:\\
  First Catalog and Rotation Periods}

\DeclareAffil{bu-iar}{Institute for Astrophysical Research, Boston University, 725
  Commonwealth Ave, Boston, MA 02215, USA}
\DeclareAffil{cfa}{Harvard-Smithsonian Center for Astrophysics, 60 Garden
  Street, Cambridge, MA 02138, USA}
\DeclareAffil{harlan}{Harlan J. Smith Fellow}
\DeclareAffil{ifa}{Institute for Astronomy, University of Hawaii at Manoa,
  Honolulu, HI 96822, USA}
\DeclareAffil{stsci}{Space Telescope Science Institute, 3700 San Martin Drive,
  Baltimore, MD 21218, USA}
\DeclareAffil{tufts}{Department of Physics and Astronomy, Tufts University,
  Medford, MA 02155, USA}
\DeclareAffil{utaustin}{The University of Texas at Austin, Department of
  Astronomy, 2515 Speedway C1400, Austin, TX 78712, USA}

\affilauthorlist{
  E. Kado-Fong\affils{tufts},
  P.~K.~G. Williams\affils{cfa},
  A.~W. Mann\affils{utaustin, bu-iar, harlan},
  E. Berger\affils{cfa},
  W.~S. Burgett\affils{ifa},
  K.~C. Chambers\affils{ifa},
  M.~E. Huber\affils{ifa},
  N. Kaiser\affils{ifa},
  R.-P. Kudritzki\affils{ifa},
  E.~A. Magnier\affils{ifa},
  A. Rest\affils{stsci},
  R.~J. Wainscoat\affils{ifa},
  C. Waters\affils{ifa}
}
\email{erin.fong@tufts.edu}

\date{\today}

\begin{abstract}
  We report on an ongoing project to investigate activity in the M~dwarf
  stellar population observed by the Pan-STARRS~1 Medium Deep Survey
  (PS1-MDS). Using a custom-built pipeline, we refine an initial sample of
  \apx4~million sources in PS1-MDS to a sample of 184,148 candidate cool stars
  using color cuts. Motivated by the well-known relationship between rotation
  and stellar activity, we use a multi-band periodogram analysis and visual
  vetting to identify 271 sources that are likely rotating M~dwarfs. We derive
  a new set of polynomials relating M~dwarf PS1 colors to fundamental stellar
  parameters and use them to estimate the masses, distances, effective
  temperatures, and bolometric luminosities of our sample. We present a
  catalog containing these values, our measured rotation periods, and
  cross-matches to other surveys. Our final sample spans periods of
  $\lesssim$1--130 days in stars with estimated effective temperatures of
  \apx2700--4000~K. Twenty-two of our sources have X-ray cross-matches, and
  they are found to be relatively X-ray bright as would be expected from
  selection effects. Our data set provides evidence that \textit{Kepler}-based
  searches have not been sensitive to very slowly-rotating stars ($\prot
  \gtrsim 70$~d), implying that the observed emergence of very slow rotators
  in studies of low-mass stars may be a systematic effect. We also see a lack
  of low-amplitude ($<$2\%) variability in objects with intermediate
  (10--40~d) rotation periods, which, considered in conjunction with other
  observational results, may be a signpost of a loss of magnetic complexity
  associated with a phase of rapid spin-down in intermediate-age M~dwarfs.
  This work represents just a first step in exploring stellar variability in
  data from the PS1-MDS and, in the farther future, LSST.
\end{abstract}

\section{Introduction}

The magnetic activity of Sun-like and lower-mass stars increases with rotation
until a ``saturation'' point is reached, past which activity becomes largely
insensitive to rotation \citep{pgr+81, vilhu1984, james2000, pizzolato2003,
reiners2009}. The details of this rotation-activity relation do not change
significantly at the transition to fully convective structure in the
lowest-mass stars \citep[$M < 0.35$~M$_\odot$;][]{chabrier2000}, and many fully
convective stars exhibit significant magnetic fields \citep{johns-krull1996,
kiraga2007,reiners2010,wdmh11}.
These are surprising
results, given that the dynamos of solar-type stars are believed to operate at
the ``tachocline,'' the transition layer between the radiative core and
convective outer envelope that, by definition, is not present
in fully convective stars \citep{c14}.

The magnetic dynamos of fully convective stars are still relatively poorly
understood. Recent theoretical efforts have worked to explain the existence of
significant magnetic activity via the effect of rotation and convective
flows \citep{dobler2006,browning2008,gastine2012,yadav2015};
because of this focus, observations of
the fully convective rotation-activity relation serve to constrain
and inform such models.

On the observational front, recent campaigns to measure rotation periods of both
fully convective and solar-type stars have benefitted greatly from
automated and semi-automated period measurement techniques
hinging upon the detection of variations in the source brightness
of time series photometry due to periodic observations of starspots on the
target star. Such studies have greatly informed our knowledge of the
distribution of stellar rotation periods with respect to stellar mass
\citep[see, e.g.][]{mcquillan2014,newton2016}; however,
systematic effects specific to each survey near the fully convective
transition make it difficult to disentangle
astrophysical trends from systematic ones.
%% \citep[][see, e.g.]{irwin2009,irwin2011,mcquillan2014,newton2016}.

Here we present measurements of the photometric rotation periods of 271
M~dwarfs in the PanSTARRS-1 Medium Deep Survey (PS1-MDS) data set, using the
multi-band Lomb-Scargle periodogram developed by \cite{vanderplas2015} to take
advantage of information provided by the five-filter photometry provided by
PS1-MDS. We additionally identify archival X-ray counterparts, which act as a
probe of the magnetic activity, for 22 of the sources.

The outline of the paper is as follows. First we discuss the PS1-MDS data set
and the construction of our source catalog and photometric database
(\autoref{dataproc}). Next we describe the identification of likely late-type
M~dwarfs in this catalog and estimates of their physical properties
(\autoref{mdwarfs}). We then present our method for identifying periodic
variations in our photometric database and verify our ability to accurately
recover rotation periods using thse methods by constructing synthetic PS1-MDS
light curves (\autoref{periodsearch}). We then present the list
of candidate rotating cool dwarfs (\autoref{catalog}).
Finally, in \autoref{discussion}, we discuss trends in our rotation period data
set against stellar mass and X-ray luminosity
in the context of recent results of contemporary studies.
We additionally examine correlations between
amplitude of variability for the final sample of rotating M~dwarfs
with respect to estimated stellar mass, as well as trends in amplitude
of variability across the five PS1-MDS filters.

\section{Data Processing}\label{dataproc}

We sought to extract light curves for all of the cool stars in the PS1-MDS
data set. To this end, we first constructed a catalog of
star-like objects in the PS1-MDS deep co-adds (\autoref{allsourcecat}), then
extracted photometry from the nightly stacks using this catalog as a reference
(\autoref{photom}). We then investigated the light curves of the objects with
colors consistent with cool stars (\autoref{mdwarfs}).

The goal in this work is to generate a small catalog of
\textit{high-confidence} cool stellar rotators, rather than a complete and/or
statistically well-characterized sample. Our general strategy in the data
processing was to set relatively loose limits on data quality in earlier
stages of the pipeline, then excise bad data farther downstream, culminating
in the visual vetting of candidate rotators (\autoref{visualvetting}).

\subsection{Observations}

The PS1-MDS was performed on the 1.8-meter PanSTARRS-1 telescope situated on
Mount Haleakala, Hawai`i, equipped with five broad-band filters,
$(grizy)_\text{P1}$, and a 1.4-gigapixel detector composed of 60 edge-abutted
4800$\times$4800 pixel CCDs with a pixel scale of 0.26~arcsec~pixel$^{-1}$
\citep{kaiser2010}. The PS1 filters are similar to those used in the Sloan
Digital Sky Survey (SDSS) but include a \yps\ filter ($\lambda_\text{eff} \sim
9600$~\AA), a bluer cutoff in the $z$ band ($\Delta\lambda \sim
1000$~\AA\ rather than 1400~\AA), a 200~\AA\ redder cutoff in the $g$ band,
and no $u$ band \citep{tonry2012}.

The Medium Deep Survey observed 10 fields spread out in right ascension over a
span of five seasons. PS1-MDS observations were conducted nightly, rotating
filters from one night to the next. Under normal conditions, \gps\ and
\rps\ were observed on a single night, with \ips\ following the next night and
\zps\ the night after, all to $5\sigma$ depths of \apx$23.3$~mag. PS1-MDS
observed \yps\ during bright periods with a $5\sigma$ depth of \apx$21.7$~mag.
The future Large Synoptic Survey Telescope (LSST) data stream will resemble
that of the PS1-MDS, but extend deeper (e.g., 5$\sigma$ single-visit depth of
24.7 in $r$~band) and significantly wider \citep[covering
  $\approx$3000~deg$^2$ per night;][]{the.lsst}.

Our analysis is based on data products from the ``PV2'' version of the PS1
data reduction. In this analysis the raw images are first processed by the
Image Processing Pipeline (IPP), which applies standard calibrations and warps
images onto a standard astrometric solution \citep{magnier2006}. As part of
our studies of transients in the PS1-MDS data set,
nightly and deep stacks of the IPP-processed PS1-MDS observations were
downloaded to Harvard University's \textit{Odyssey} high-performance computing
cluster and ingested into the \textsf{photpipe} pipeline originally developed
for the SuperMACHO and ESSENCE projects \citep{rest2005, rsf+14}. We conducted
our study at a time when Pan-STARRS project resources were dedicated to the
development of the ``PV3'' public data release and so we chose to derive our
source catalogs and extract photometry using the resources available locally,
namely the nightly and deep PV2 stack images. We did so using customized
routines operating alongside the \textsf{photpipe} framework as described
below. Future iterations of this project will leverage the final PV3 data
products.

\subsection{All-Source Catalog}
\label{allsourcecat}

We generated source catalogs by running \textsf{SExtractor} \citep{bertin1996}
on the \textsf{photpipe}-ingested \textsf{notyr1} stack images, which combine
all observations except for those obtained during the first year of PS1-MDS
observations. Routines in \textsf{photpipe} normalize the
\textsf{SExtractor}-reported fluxes onto the absolute photometric system
defined by the \textsf{pv2e} ubercalibration of PS1 (D.~Finkbeiner, 2016,
priv. comm.; see also \citealp{fss+16}). Because the stack images are of high
quality, we used a low source detection threshold of $0.1\sigma$ above the sky
level (\textsf{SExtractor} parameters \textsf{DETECT\_THRESH} and
\textsf{ANALYSIS\_THRESH}), then flagged as dubious sources that had
implausible \textsf{SExtractor}-reported parameters. In particular, we flagged
sources that did not meet the following criteria:
\begin{itemize}
\item magnitude $2 <$ \textsf{m} $< 28$~mag,
\item magnitude error \textsf{dm} $< 0.4$~mag,
\item no flag indicating nearby neighbors or bad pixels,
\item no flag indicating that the object was deblended,
\item major-axis FWHM $0 <$ \textsf{fwhm1} $< 25$~pixels,
\item minor-axis FWHM $0 <$ \textsf{fwhm2} $< 185$~pixels,
\item background level $900 <$ \textsf{sky} $< 1200$ ($griz_\text{P1}$ images)
  or $2600 <$ \textsf{sky} $< 4000$ (\yps).
\end{itemize}
These cuts were determined empirically by examining the distributions of the
parameters reported by \textsf{SExtractor}. The bimodality in the last
criterion was needed because the \yps\ band images are normalized to a
different background level than images using the other filters. We further
identified likely stellar sources using the following criteria:
\begin{itemize}
\item star/galaxy classifier \textsf{class} $\ge 0.9$,
\item \textsf{elongation} $< 3$.
\end{itemize}
In practice, the constraint on \textsf{elongation} eliminated only a handful
of candidate stellar sources in which \textsf{SExtractor}'s
neural-network--based star-galaxy classifier misbehaved.

We constructed a final catalog of sources by merging the \textsf{SExtractor}
source lists generated in the five filters with the list of photometric
standards associated with the \textsf{pv2e} calibration data set, using a
0.3$''$ positional match tolerance. We additionally flagged any sources that
were detected in fewer than three of the five filters. The final source
catalog contains 4,073,661 likely stellar sources not flagged by the above
criteria. This may be compared to 342,762 objects in the \textsf{pv2e}
catalog, which is limited to moderately bright sources with excellent
photometric properties.

\begin{deluxetable}{lcc}
  % XXX: would be good to generate this table programmatically to make
  % sure that the numbers stay up-to-date!
  \tablewidth{0em}
  \tablecaption{Stellar locus slopes\label{locusparams}}
  \tablehead{
    \colhead{Color-color plane} &
    \colhead{Slope} &
  }
  \startdata
  \gri\tablenotemark{a} & $0.403$ \\
  \riz & $0.449$ \\
  \izy & $0.417$
  \enddata
  \tablenotetext{a}{This locus transitions to be vertical at $(g-r, r-i)_\text{P1} =
    (1.23, 0.55)$~mag; see \autoref{colormasks}.}
\end{deluxetable}

\subsection{Photometry}
\label{photom}

We extracted photometry from the nightly stacks using \textsf{SExtractor} in
conjunction with \textsf{PSFEx} \citep{b11b} to perform both PSF fitting and
aperture photometry. Here \textsf{SExtractor} was run in ``double-image'' mode
using the \textsf{notyr1} deep stacks to detect sources and the per-night
images to measure their photometry; this stategy makes it possible to detect
flares from stars that are seen in the deep stacks but generally too faint to
be detected in the nightly stacks. We calibrated photometry for the individual
epochs to the absolute scale by matching to the \textsf{pv2e} catalog and
solving for a scale factor to apply to the measurements, $\log f_\text{abs} =
\log_\text{psf} + m$. We used Markov Chain Monte Carlo (MCMC) sampling with
\textsf{emcee} \citep{fmhlg13} to determine both the scale factor $m$ and its
uncertainty while accounting for the measurement uncertainties present in both
catalogs. Epochs were rejected where the reduced $\chi^2$ of the
absolutization fit exceeded 2, the scale factor was negative, the fractional
uncertainty in the scale factor exceeded 25\%, or fewer than 10 sources were
identified. (The typical 2960$\times$2960 pixel$^2$ [740$\times$740
  arcsec$^2$] image subsections that we analyze have \apx2,000 sources, but
many fewer can be present when the weather was poor or the subsection is on
the very edge of the night's sky coverage.) The final photometric data set
consists of 443,747,494 measurements and upper limits across the five PS1-MDS
filters, with \apx41~million in \gps, \apx72~million in \rps, \apx149~million
in \ips, \apx140~million in \zps, and \apx43~million in \yps.

We obtained mean photometry for each of the cataloged sources by applying the
same technique to the \textsf{notyr1} stacks. In these deeper images we found
that we needed to add a fractional 3\% photometric uncertainty in quadrature
to the pipeline-reported values to ensure that the reduced $\chi^2$ of the
typical epoch stayed within the limit specified above. We also identified a
handful of cases where the absolutization fit for the PSF fluxes required an
additional term: $\log f_\text{abs} = k \log_\text{psf} + m$, where by default
it was implicitly the case that $k = 1$. The median \zps\ magnitude in our
catalog is 23.31~mag, compared to 18.72~mag in \textsf{pv2e}.

\begin{figure*}
  \includegraphics[width=\linewidth]{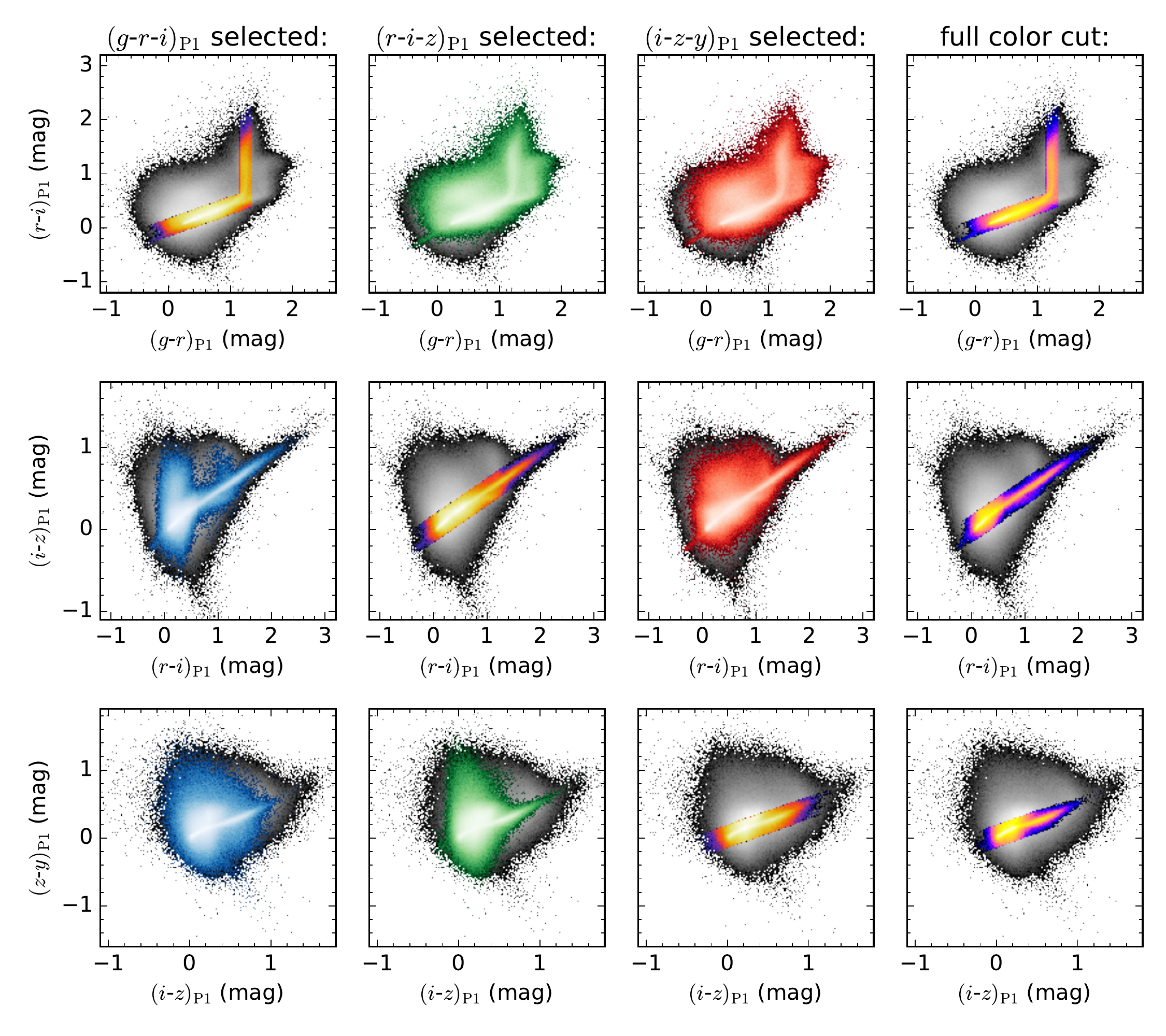}
  \caption{Color cuts applied to isolate the stellar locus in each color-color
    space. In each panel, the overall catalogued sample is shown in a
    grayscale 2D histogram, and selected sources are shown in color. In all
    cases the intensity scaling is logarithmic. \textit{Leftmost column:} The
    results of the stellar locus cut in \gri\ plane. The top panel shows the
    distribution of sources that pass the \gri\ cut in the \gri\ color-color
    plane. The middle panel shows the \gri-selected sources in the
    \riz\ plane, and the lower panel shows them in the \izy\ plane.
    \textit{Second column:} Similar to the leftmost column, but showing the
    results of the \riz\ cut in the three planes. \textit{Third column:}
    Similar to the others, but showing the results of the \izy\ cut.
    \textit{Rightmost column:} For each row, the result of the full color cut
    (the intersection of the \gri, \riz, and \izy ~individual color cuts)
    applied to \gri, \riz, and \izy ~color-color spaces, respectively.}
  \label{colormasks}
\end{figure*}

\subsection{Candidate Cool Dwarfs}\label{mdwarfs}

We identified candidate cool dwarfs using color cuts as described below.
Because the PS1 filters are not precisely the same as those used by SDSS,
especially on the red end that is of greatest interest to us, we derived these
cuts independently rather then reusing prior SDSS-based results
\citep[e.g.,][]{kowalski2009}, although alternatively PS1-SDSS color
transformations could be applied \citep{tonry2012}.

\autoref{colormasks} shows color-color diagrams for our flagged source
catalog. In each such plane stellar sources are generally confined to a
``stellar locus'', with the surrounding sources generally corresponding to
galaxies \citep[e.g.,][]{mga+15}. We note that the reddening vector is fairly
well-aligned with the stellar locus in the \riz\ and \izy\ planes
\citep{gsf+15} and so reddening should not scatter many cool star candidates
off the locus. We show below that for the stars that are bright enough for us
to detect their periodic variability, reddening is generally a small effect.

For each color-color plane we determined the location of the stellar locus by
binning the catalogued source colors into a 400$\times$400 two-dimensional
histogram, then fitting a slope to the positions of the 1600 most-populated
bins. In the case of the \gri\ space, we modeled the locus as discontinuously
transitioning to be vertical at $(g-r, r-i)_\text{P1} = (1.23, 0.55)$~mag as
shown in \autoref{colormasks}. This transition point marks the approximate
beginning of the M~dwarf sequence \citep[e.g.,][]{kh07, mga+15}. The
parameters derived from these fits are reported in \autoref{locusparams}.

\autoref{colormasks} shows the results of the stellar locus color cut in each
color-color space in addition to their intersection. As shown in the
off-diagonal panels of \autoref{colormasks}, the sources identified by a
stellar locus color cut in any given color-color plane are subject to a large
amount of scatter in the other two planes. The combination of the locus color
cuts is therefore crucial in identifying the true stars in the sample.
Although sources with upper limits consistent with the above criteria are
included in the final sample, they are not shown in the figure.

We identified sources as potential cool stars if they were likely stellar
sources (cf. \autoref{allsourcecat}) with mean catalog photometry meeting the
following criteria:
\begin{itemize}
\item detections in the \ips\ and \zps\ bands,
\item position consistent with the stellar locus,
\item $(g-r)_\text{P1} > 1.0$~mag,
\item $(r-i)_\text{P1} > 0.55$~mag, and
\item $(i-z)_\text{P1} > 0.32$~mag.
\end{itemize}
The latter criteria roughly isolate stars of M spectral types and later. Here,
based on inspection of color-color histograms, consistency with the stellar
locus is defined as colors consistent with lying within 0.15~mag of our locus
fits in all three color-color spaces. Sources near the vertical portion of the
\gri\ locus had to lie within 0.10~mag of it. The total number of sources
meeting these criteria was 184,148.

\section{Search for Periodicities}\label{periodsearch}

We searched for periodic variations in the light curves of our candidate cool
stars using the multiband extension of the Lomb-Scargle periodogram introduced
by \citet{vanderplas2015}. In this method, each source's light curve is
modeled as a combination of a ``base'' term, common to every filter, and
``band'' terms unique to each observed filter. Each term is expressed as a
truncated Fourier series with $N_\text{base}$ and $N_\text{band}$ terms,
respectively. This configuration yields two fundamental models, the
``shared-phase'' model ($N_\text{base} = 1$, $N_\text{band} = 0$) and the
``multi-phase'' model ($N_\text{base} = 0$, $N_\text{band} = 1$), as termed by
\cite{vanderplas2015}. For each candidate periodicity, the model is optimized
in a least-squares sense using a Tikhonov regularization to push as much
variation into the ``base'' term as the data allow.

In the case of starspots passing through the view of the
observer, we expect the filters to share the same phase;
we find that for the purposes of this study,
the best results were achieved by computing periodograms with the shared-phase
model after normalizing the light curves of each filter by dividing out their
mean values. Though periods can still be recovered accurately without
normalization, we choose to normalize the light curves because
the amplitudes of unnormalized
light curves tend to vary significantly from filter to filter.
Our experience is consistent with various other studies in that
extending past the simple sinusoidal model (i.e., $N_\text{base} > 1$) is not
required for accurate period determination \citep[see, e.g., ][]{newton2016}.

\begin{figure}
  \includegraphics[width=\linewidth]{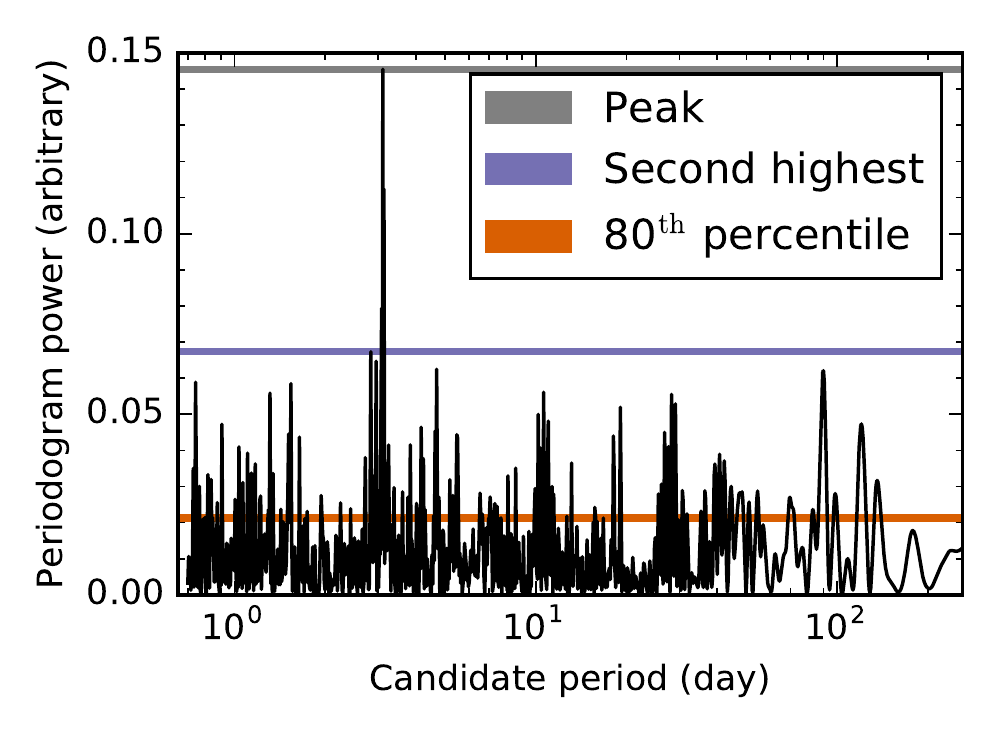}
  \caption{An example periodogram. \pr{} is calculated by dividing the power
    of the best-fit peak (shown in grey) and the power of the second-highest peak
    (shown in purple). \paeight\ is calculated by dividing the power of the
    best-fit peak and the $80^{th}$ percentile power value (shown in orange).}
  \label{pgramexample}
\end{figure}

For each candidate cool star, we computed a periodogram using the
\textsf{gatspy} implementation of the multiband periodogram provided by
\citet{vanderplas2015}, sampling 1000 periods logarithmically spaced between
0.7 and 300 days. We identified the best-fit period as the one resulting in
the highest periodogram power, ignoring peaks between 0.9 and 1.1 days to
avoid false signals associated with the observing cadence. We computed two
significance metrics. \pr{} denotes the ratio between the powers of
the best-fit and second-highest peaks in the periodogram, once again ignoring
peaks around 1~day, while \paeight{} refers to the ratio between the power at the
best-fit peak and the 80th percentile periodogram power. These parameters are
illustrated in \autoref{pgramexample}. We use cuts on these significance
parameters and visual vetting to identify rotating objects, which we describe
below.

\begin{figure*}[tb]
  \includegraphics[width=\linewidth]{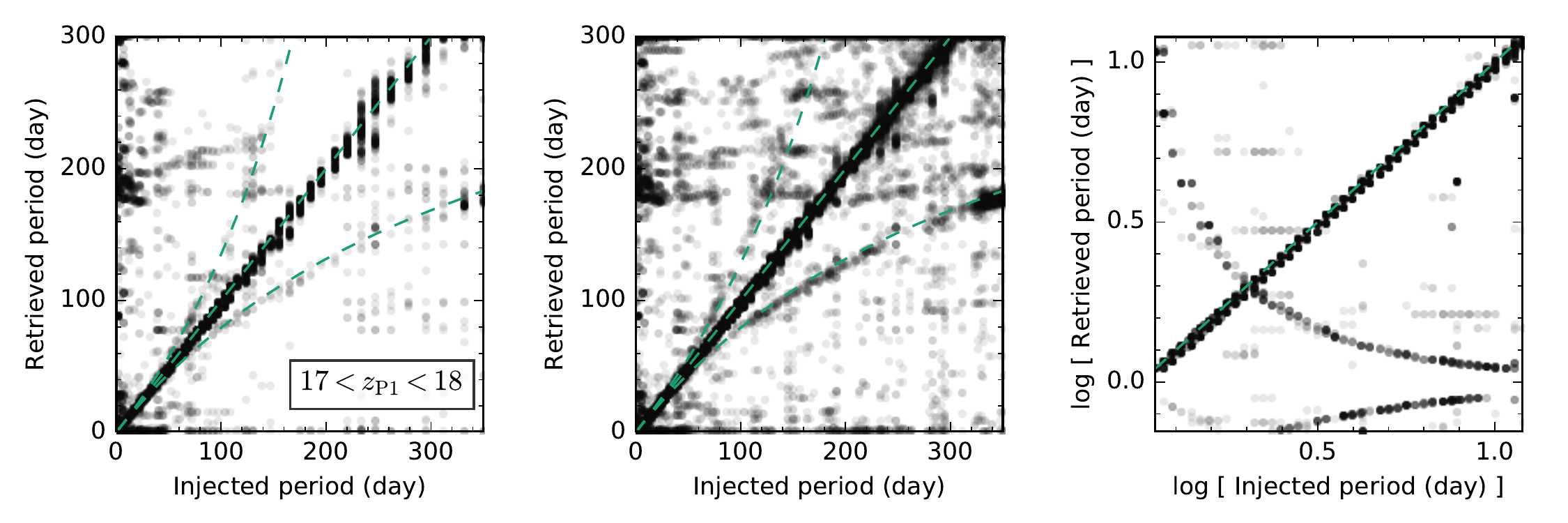}
  \caption{Input period versus output period for the synthetic light curves.
    \textit{Left panel:} Input period
    versus output period for only those sources with $17 < \zps < 18$.
    \textit{Middle panel:} The same, for all sources in the synthetic catalog
    ($\zps < 22.$). \textit{Right panel:} A log-scale zoom in of the middle
    panel for injected periods between 0.7 and 20 days. Beat frequencies with
    the one day observing cadence are clearly visible along the curved tracks,
    as shown by the dashed green lines.
  }
  \label{pinpout}
\end{figure*}

\begin{figure}[tb]
  \includegraphics[width=\linewidth]{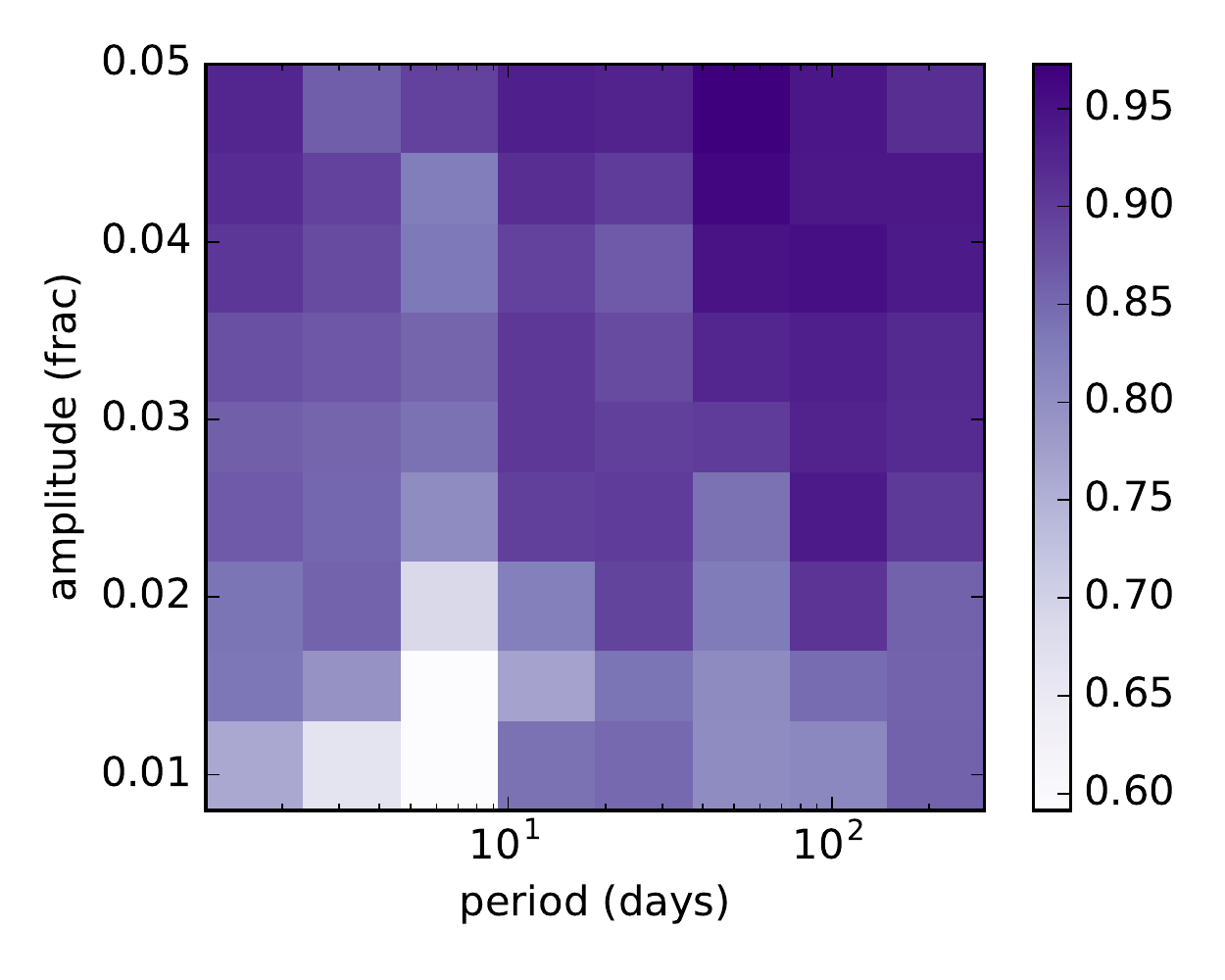}
  \caption{Period recovery fraction as a function of injected period (x-axis)
    and amplitude (y-axis). Synthetic light curves with a low amplitude of
    variability and short period are recovered relatively less often than the
    bulk of the synthetic sample. We attribute the low success rate at $3
    \lesssim (\prot / \text{day}) \lesssim 10$ and small amplitudes to the effect
    of aliases with with observing cadence and 
    harmonics (\autoref{pinpout}, rightmost panel).
    Injected periods outside the recovery range (i.e. $\text{P}_\text{rot} >
    300$~days) are excluded from this figure. Each bin contains between 215
    and 680 contributing sources, with a median of 480 contributing sources.}
  \label{synth_pcolor}
\end{figure}

\subsection{Synthetic Light Curves}\label{syntheticlightcurves}

To calibrate our significance cuts and test the robustness of the periodogram
analysis, we computed periodograms for light curves with artificial
periodicities, injecting sinusoidal signals into the light curves of sources
that had the lowest variability significance ratings ($\pr < 1.01$, where 1.0
is the lowest possible value of \pr). We drew quiescent light curves from
across the survey at random in order to minimize the impact of
location-dependent features of the light curves (due to, e.g. differences in
the total number of observations). The injected signals had periods of
0.1--350~days and amplitudes of 1--5\% that of the mean flux in each filter.
We set the range of injected periods outside the domain of the periodogram
search (0.7--300~days) in order to investigate the behavior of the periodogram
when the true rotation period is outside the range of possible periods.

We generated a set of 30,884 synthetic lightcurves using quiescent sources
with $17.0 < \zps < 18.0$~mag. This range was chosen based upon the apparent
magnitudes of the rotating M~dwarfs in our final sample. The overall rate of
successful recoveries was 79.5\%, where we define a successful recovery as one
in which the the strongest periodogram peak met our significance cuts (see
below) and the periodogram-derived period matched the injected period to
within 30\%. For 9.5\% of sources, an accurate period (again, where the derived and
injected periods matched to within 30\%) was recovered, but the periodogram
did not pass our significance cuts.
We generated an additional set of synthetic light curves from
sources with $\zps < 22$~mag in order to test recovery rates for the dataset
as a whole. In the $\zps < 22$~mag sample, the successful recovery rate was
73.8\%. 12.2\% of sources in this sample had accurate periods with periodograms
that did not pass our signficance cuts.

The leftmost panel of \autoref{pinpout} plots the injected period versus the
recovered period in the bright sample of synthetic light curves, while the
middle panel shows the same for the union of the two synthetic samples.
Although aliased periods are visible along curved tracks about the 1:1 line in
both panels, the aliased periods (and other incorrectly recovered periods) are
significantly more prominent in the full data set. Beat frequencies between
the true period and one-day observing cadence can also be seen at small
periods, as shown in the rightmost panel of \autoref{pinpout}.

\autoref{synth_pcolor} shows the recovery fraction as a function of injected
period and amplitude of variability for the full synthetic data set. A drop in
the recovery fraction between 2 and 10 days is visible, which we attribute to
the beat frequencies shown at right in \autoref{pinpout}. The periodogram
shows a moderate increase in performance with increasing period at amplitudes
above 2.0\%. At amplitudes below 2.0\%, there is no strong increase in
recovery fraction with period.

We additionally tested our ability to recover periodic signals faster than the
daily observation cadence. We generated a sample of 4112 synthetic light
curves with injected periods ranging from 0.1 to 1.0~days. The shortest
injected period that was successfully recovered was 0.613~days. For periods
between 0.613 and 1.0~days, the successful recovery rate was 36\%,
significantly lower than the recovery rate for periods above 1~day. However,
of the 1307 sources with a recovered period of ${<}1.0$ day, 72.5\% of sources
had an injected period of ${<}1.0$ day.

\begin{deluxetable}{lr}
  \tablewidth{0in}
  \tablecaption{Successful and unsuccessful recoveries in synthetic
    data\label{synthtable}}
  \tablehead{
    \colhead{Cut type} &
    \colhead{Successes : failures}
  }
  \startdata
  $\text{\paeight} > 10.5$ & 38119 : 2548 \\
  $\text{\pr{}} > 2.3$ & 6211 : 196 \\
  $\text{\paeight} > 10.5$ and $\text{\pr} > 2.3$ & 6074 : 176  \\
  $\text{\paeight} > 10.5$ or $\text{\pr} > 2.3$ & 38256 : 2568  \\
  \enddata
\end{deluxetable}

\begin{figure*}[ht]
  \includegraphics[width=\linewidth]{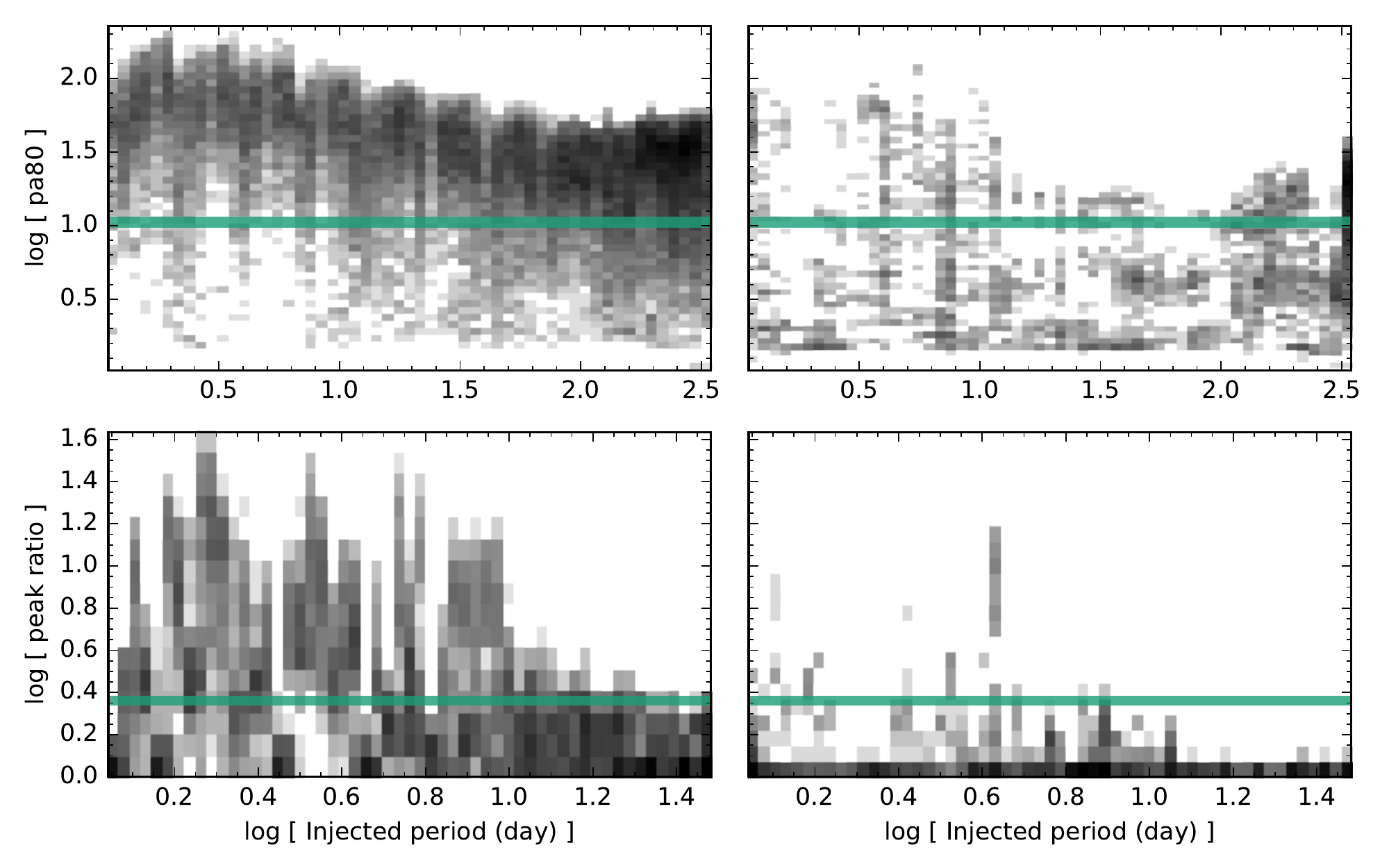}
  \caption{\textit{Left column:} The distribution of input period versus
    \paeight\ (\textit{top row}) and \pr{} (\textit{bottom row}) for those
    sources in which the recovered period is within 30\% of the true period.
    \textit{Right column:} The same for those sources in which the recovered
    period is not within 30\% of the true period. The green horizontal lines
    show the cutoffs used to select sources for visual vetting.}
  \label{prvpa}
\end{figure*}

\subsection{Significance Cuts and Visual Vetting}
\label{visualvetting}

We chose a subsample of periodograms to vet visually using significance cuts
on the \paeight and \pr\ metrics. \autoref{prvpa} shows the distributions
these values in the synthetic sample. We examined all periodograms having
$\paeight > 10.5$ or $\pr > 2.3$, as shown by the green lines. While \pr\ is
not a useful significance metric for periods above \apx10~days, we included it
in the significance cut because there are 137 synthetic rapid rotators that
pass the \pr{} cut but not the \paeight\ cut. The overall numbers of
successful and unsuccessful recoveries yielded by these cuts in the full
synthetic data set are presented in \autoref{synthtable}.

Concerning rapid rotators, 558 sources with an injected period of less than
1.0 day had a strongest periodogram peak located at $P \leq 1$~day that passed
our significance cuts. Conversely, 117 sources with an injected period of more
than 1.0 day were classified as rotators with a period of less than 1.0 day.
We therefore conclude that the periodogram is able to identify sources with
rotation periods faster than the cadence of observation, but does not
constrain their periods well.

In the full sample of candidate cool stars, 1,067 sources met the significance
criteria. Two authors (EKF and PKGW) then separately performed a visual
vetting of these sources' light curves, periodograms, and SEDs. Of the 1,067
candidates, both vetters independently agreed that 271 of them were secure
detections of rotation.

The uncertainty in the measured rotation period increases with increasing
period, given that fewer periods are observed with the time frame of the
survey. To estimate this uncertainty, we ran the multiband Lomb-Scargle
periodogram on the final sample of rotators with a more finely spaced array of
trial rotation periods, and measure the width of the primary peak of the
periodogram. We caution, however, that the sources with $P_\text{rot} <
1.0$~days are subject to greater uncertainty than the value derived in this
way, for the reasons described above.

\section{A Catalog of PS1 Rotating Cool Dwarfs} \label{catalog}

\begin{figure*}
  \includegraphics[width=\linewidth]{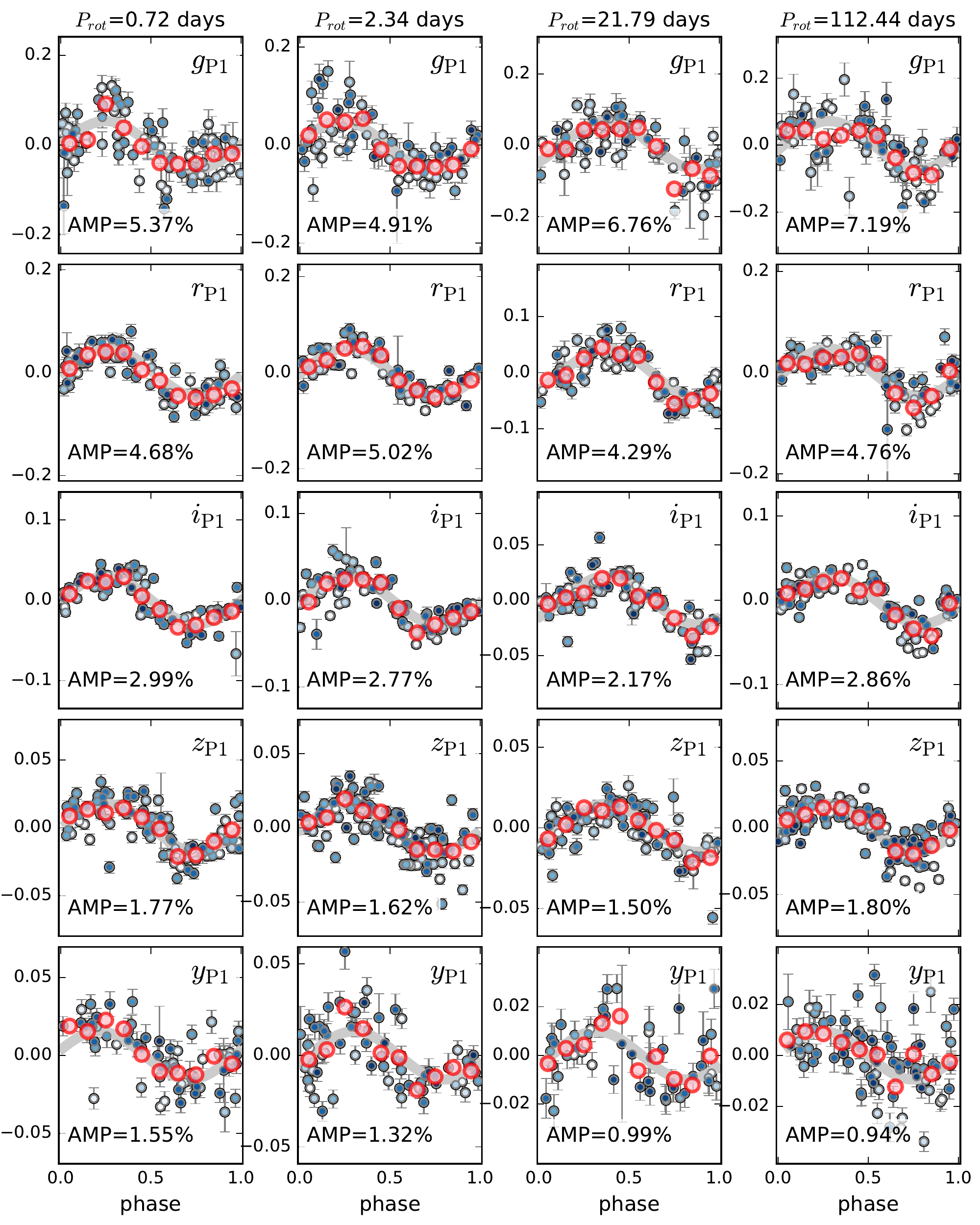}
  \caption{Phased light curves drawn from the final sample of rotating
    M~dwarfs, with periods spanning 0.7 to 113 days. The red points shows the
    phased light curve of the source, colored by time of observation, the
    purple points show the binned mean flux, and the model of the light curve
    for each filter is shown by the gray curve.
    The amplitude of variability estimated from the model of the source light
    curve is tabulated in the bottom left corner of each panel.
    From left to right, the
    sources plotted are: \object{WISEA~J104946.22$+$573026.7},
    \object{WISEA~J084921.27$+$444949.2},
    \object{WISEA~J100031.55$+$032820.9}, and
    \object{WISEA~J083701.66$+$441542.6}.}
  \label{samplelightcurves}
\end{figure*}

We list the confirmed rotators in \autoref{t.rotators}. A sample of phased
light curves with periods sapnning 0.72--112.4 days 
is shown in \autoref{samplelightcurves}. We examined image
cutouts of these sources in the MDS deep stacks, 2MASS Atlas images
\citep{the.2mass}, and the ``unWISE'' unblurred WISE coadds including
NEOWISE-Reactivation data \citep{l14b, mls16}. Nearly all (265) of the
confirmed rotators are securely detected in the AllWISE source catalog
\citep{cwc+13}, and we use these identifiers whenever possible. Five
additional sources are detected in the Sloan Digital Sky Survey Data Release~9
catalog \citep{the.sdssdr9}. Fewer sources are detected in 2MASS (230) than
either of these surveys. One source is not found in any of the standard
all-sky catalogs we have consulted, and we name it
\object{PSO~J053.3285$-$27.1683}; it lies outside the SDSS footprint. In every
case, the sources without corresponding AllWISE catalog entries are
perceptible in the unWISE coadds, but are blended with brighter neighbors due
to the telescope's relatively poor 6$''$ resolution. This suggests that
improvements in the AllWISE source extraction pipeline could lead to their
recovery and measurement of their WISE photometry.

\floattable
\input{rotators-sample}

\autoref{t.rotators} includes dereddened \zps\ magnitudes (see below) and
$(grizy)_\text{P1}$ colors derived from the \textsf{notyr1} stack images,
placed onto the ubercalibrated \textsf{pv2e} absolute photometric system as
described in \autoref{dataproc}. We additionally report estimated effective
temperatures, masses, distances, and bolometric luminosities computed as
described below. While our effective temperature measurements are believed to
be fairly accurate (\apx3\%), the other quantities are difficult to determine
from colors alone and are uncertain at the \apx20\% level. Finally, we report
the key parameters determined from our periodicity search
(\autoref{periodsearch}).

\subsection{Estimated Stellar Properties}
\label{stellar_properties_summary}

We estimated the physical properties of the vetted rotators from their colors
and apparent magnitudes. We describe the full procedure in
\autoref{stellar_properties} but also summarize it here. First, using the
filter bandpasses provided by \citet{tonry2012} and the flux-calibrated
spectra and fundamental parameters of the M~dwarf sample developed by
\citet{mann2015, mfg+16.erratum}, we computed new polynomial relations between
PS1 colors and several stellar parameters: mass ($M$), effective temperature
(\teff), absolute \zps\ magnitude ($M_{z,\text{P1}}$), and the \zps-band
bolometric correction factor ($\text{BC}_{z,\text{P1}}$).
\autoref{stellar_properties} includes tables of the polynomial coefficients
for use in future PS1 studies of cool stars.

We then simultaneously estimated the distance and reddening of each star,
combining our $M_{z,\text{P1}}$ relation with the three-dimensional PS1 dust
maps of \citet{gsf+15}. In the final catalog the median $A_V$ is 0.04~mag, the
95th percentile is 0.22~mag, and the typical distance uncertainty is 20\%.

Finally, we used our polynomial relations and dereddened colors to estimate
\teff, $M$, and the bolometric flux $f_\text{bol}$ for each rotator. A
detailed description of the derivations of each of these parameters may be
found in \citet{mann2015, mfg+16.erratum}. Based on the scatter in our
polynomial fits and the \teff\ calibration standards used by \citet{mann2015,
  mfg+16.erratum}, the uncertainties in our \teff\ values are 80~K, or
$\lesssim$3\%, and those on $f_\text{bol}$ are \apx4\%. The uncertainties in
$M$, on the other hand, are \apx20\%.

\subsection{Cross-Identifications}

We cross-matched our catalog against the SDSS DR9 catalog \citep{the.sdssdr9},
the AllWISE catalog \citep{cwc+13}, the 2MASS Point Source Catalog
\citep{the.2mass}, the UKIDSS Large Area Survey catalog \citep{lwa+07}, and
the third (2005 September) release of the Deep Near-IR Survey of the Southern
Sky \citep[DENIS;][]{vpb+99} using the Vizier web service \citep{obm00}
through the \textsf{astroquery} Python
module\footnote{\url{http://www.astropy.org/astroquery/}}.
\autoref{crossmatch_summary} summarizes the results of the cross-matches and
stands in for a machine-readable table (MRT) provided with this article that
includes the cross-matched identifiers and basic photometric measurements in
the 19 additional filters provided by these surveys.

\begin{deluxetable}{lrl}
  \tablewidth{0in}
  \tablecaption{Summary of MRT of optical/IR cross-matches\label{crossmatch_summary}\tablenotemark{a}}
  \tablehead{
    \colhead{Survey} &
    \colhead{Number of matches} &
    \colhead{Photometric filters}
  }
  \startdata
  This work & 271\tablenotemark{b} & $(grizy)_\text{P1}$ \\
  SDSS & 249 & \textit{u g r i z} \\
  AllWISE & 265 & W1 W2 W3 W4 \\
  2MASS & 230 & J H K$_s$ \\
  ULAS & 81 & Y J H K \\
  DENIS & 64 & I J K
  \enddata
  \tablenotetext{a}{This table summarizes the contents of a machine-readable
    table containing cross-matched optical/IR identifiers and photometry.}
  \tablenotetext{b}{For convenience, the MRT duplicates the
    $(grizy)_\text{P1}$ photometry presented in \autoref{t.rotators}.}
\end{deluxetable}

We also cross-matched our catalog with a variety of X-ray surveys. After
manual inspection of the cross-match results, we identified X-ray counterparts
for 22 sources. The X-ray properties of these sources are summarized in
\autoref{t.xray}. We have preferred measurements from the \textit{Chandra}
Source Catalog (CSC) version 1.1 \citep{epg+10} when available (eight
sources). Eleven additional sources have matches in the 3XMM-DR5 catalog
\citep{rww+16}, and the final three sources are found in the XMM-LSS catalog
\citep{ccp+13}. In \autoref{t.xray} we have converted X-ray fluxes to a common
energy band of 0.2--2~keV as in \citet{cwb14} from the energy ranges used in
the catalogs: 0.5--7~keV for CSC~1.1; 0.5--4.5~keV for 3XMM-DR5; and
0.5--2~keV for XMM-LSS. Using PIMMS and an APEC model to determine the
conversion factors, we multiplied the catalog fluxes by 1.21, 1.21, and 1.26,
respectively (see \citealt{cwb14} for details).

\subsection{Contaminants}
\label{contaminants}

Our optically-based selection of cool stars could result in contamination of
the sample by M~giants. Lacking spectroscopic follow-up, some insight into the
contamination can be gained by considering infrared colors \citep{bb88,
  lsz+16}. For those of our sources with robust cross-matches to both the
2MASS and WISE catalogs, we have tested the criterion given in Equation~1 of
\citet{lsz+16}, which categorizes probable M~giants based on their position in
the W1$-$W2/$J{-}K$ color space. Here we did not attempt to deredden the
catalogued 2MASS and WISE magnitudes. We find three sources that are
classified as probable giants by the criterion:
\object{WISEA~J022325.43$-$052529.4} ($\prot = 36.2 \pm 0.3$~d),
\object{WISEA~J095821.50$+$030242.6} ($\prot = 16.30 \pm 0.05$~d), and
\object{WISEA~J221306.75$-$001313.0} ($\prot = 30.8 \pm 0.2$~d). While 101
sources lack sufficient information to evaluate the criterion, examination of
our catalog in other color spaces such as $(g{-}i)_\text{P1}$/$J{-}K$
\citep{blee11} does not yield any likely examples of additional giant
contaminants. \autoref{t.rotators} flags the three objects selected as
probable giants. They are not included in subsequent analysis.

By construction, any spatially unresolved binaries in our catalog must have
$(g{-}r{-}i{-}z{-}y)_\text{P1}$ colors consistent with single M~stars. Since
the hotter star in such a pair dominates the radiative output, such systems
likely contain two low-mass stars and should not be considered
``contaminants'' \textit{per se}. However, our estimates of mass and other
stellar parameters will be inaccurate in such systems. Lacking spectroscopic
follow-up or precise distance estimates, we are unable to identify probable
binaries in the current data set. \citet{cal+16} identified a sample of 132
young, low-mass ($M < 0.5$~\msun) stars with photometric rotation periods in
the Pleiades. Leveraging the fact that their targets all lie at approximately
the same distance, they identified 20 of these sources as likely binaries from
their position on a $V{-}K$ color-magnitude diagram. If the multiplicity in
this cluster is approximately the same as that in the field, this suggests
that \apx15\% of our sources are unresolved binaries

\begin{figure}[tb]
  \includegraphics[width=\linewidth]{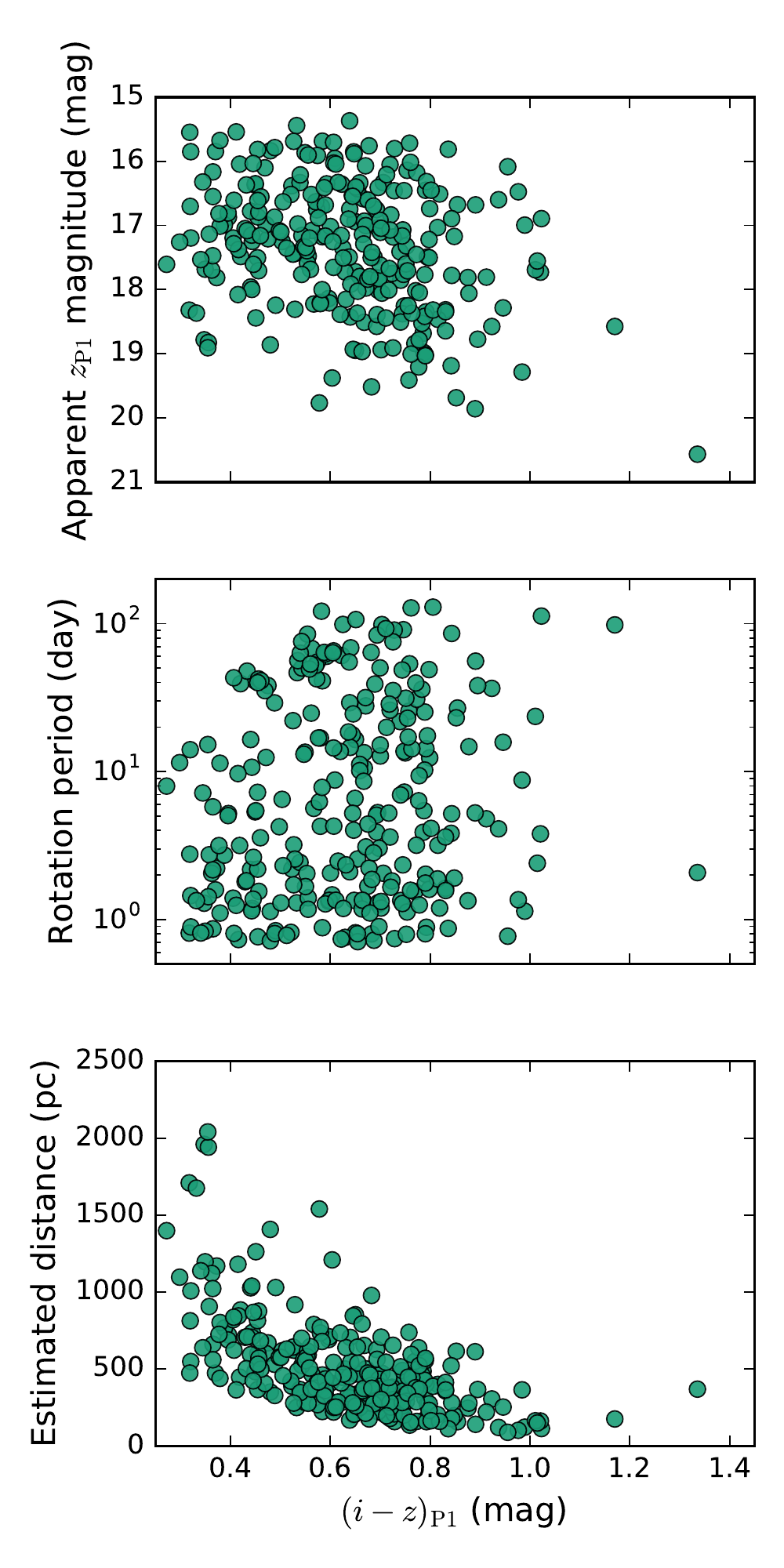}
  \caption{Observed and derived properties of our sample of 271 rotating cool
    stars. From top to bottom: apparent \zps\ magnitude, rotation period, and
    estimated distance (\autoref{stellar_properties}).}
  \label{sample}
\end{figure}

\begin{figure*}[tb]
  \includegraphics[width=\linewidth]{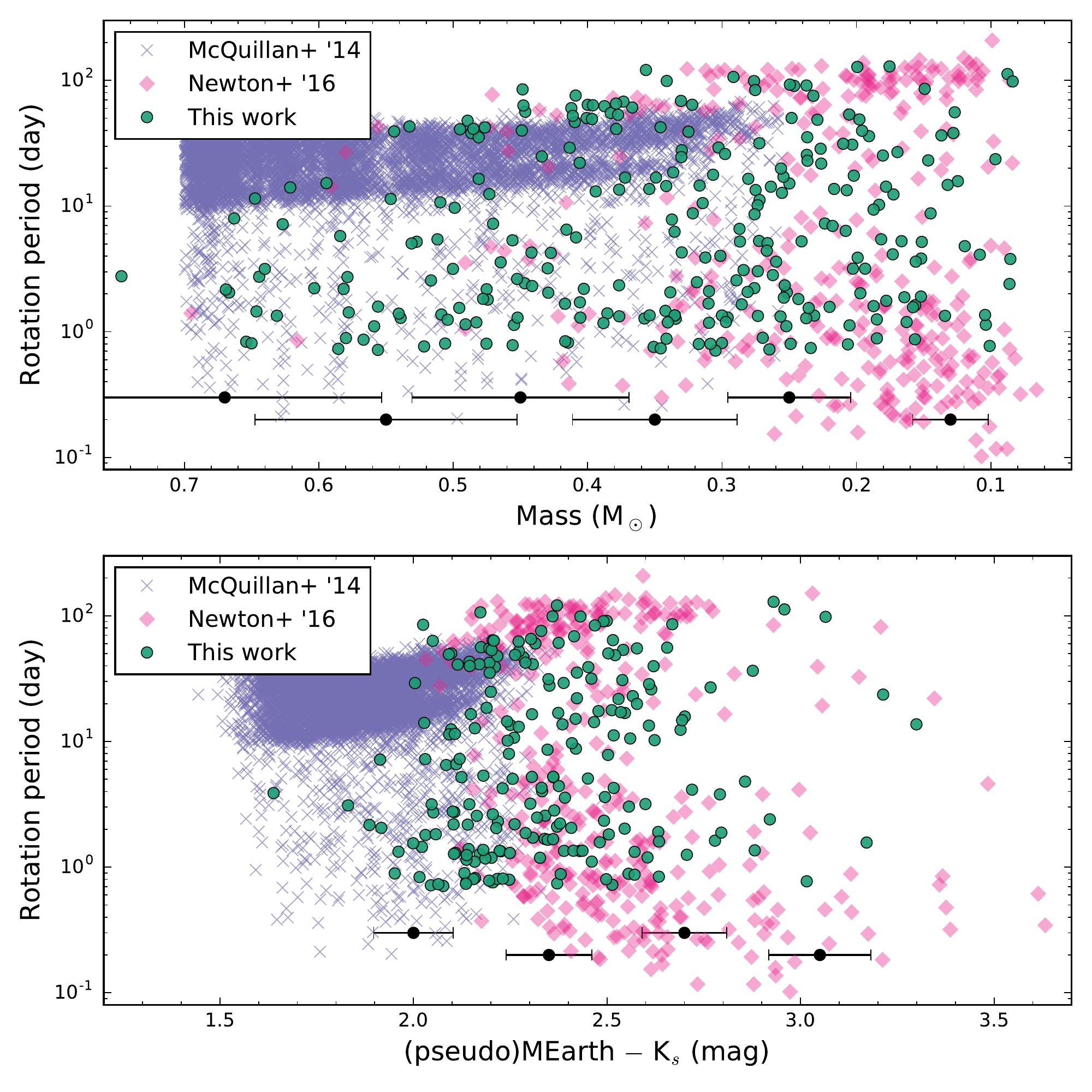}
  \caption{Three samples of stars with photometric rotation measurements.
    \textit{Purple x's}: \apx6,200 stars from an overall sample of \apx34,000
    found in the \textit{Kepler} data set \citep{mcquillan2013,
      mcquillan2014}. \textit{Pink diamonds}: 428 stars found in the MEarth
    data set \citep{newton2016}. \textit{Green circles}: 271 stars found in
    this work. \textit{Black errorbars} show typical uncertainties in the
    abscissa values for our data. \textit{Upper panel}: the full sample, using
    estimated stellar masses as the abscissa. Each study estimates masses
    using a different method, and the estimates used by \citet{mcquillan2013,
      mcquillan2014} are known to be biased \citep{dc13}. \textit{Lower
      panel}: a reduced sample using $\text{(pseudo)Mearth} - K_s$ as the
    abscissa, discarding non-MEarth sources without 2MASS cross-matches
    \citep[see \autoref{discussion} and][]{dicn16}. Our sample bridges the
    mass/color ranges best probed by \textit{Kepler} and MEarth. We emphasize
    that each study is subject to different selection effects, so
    intercomparisons should be performed with care.}
  \label{mass-prot}
\end{figure*}

\section{Discussion} \label{discussion}

\subsection{Characteristics of the Rotating Sample}

\autoref{sample} visually summarizes the characteristics of the stars with
rotation periods detected in our study. The $(i - z)_\text{P1}$ colors of the
bulk of the stars range from \apx0.3--1.1~mag, corresponding to a mass range
of \apx0.7--0.09~\msun\ and a \teff\ range of \apx3900--2800~K in our adopted
models. Two unusally red sources have $(i - z)_\text{P1} = 1.14$
(WISEA~J160316.52$+$541556.7, $\prot \sim 100$~d) and $(i - z)_\text{P1} =
1.38$ (WISEA~J221445.29$+$004500.7, $\prot \sim 2$~d). The detected rotation
periods range between \apx1--130~d, with an approximately uniform distribution
in $\log \prot$. Although our tests with synthetic PS1-MDS light curves
indicate that we should be sensitive to extremely slow ($\prot > 150$~d)
rotators, we do not find such a population in the present sample. However, it
is important to note that the sample at hand includes only the most secure
period measurements of the dataset --- 40 out of 1067 of the sources in the
automated sample did have $\prot > 150$, but were removed from the final
sample during visual vetting. We therefore cannot exclude the possibility
that there exist very slow rotators in the data that were ruled as marginal
detections during the construction of the present sample. The median estimated
distance is 440~pc, with the redder (and therefore intrinsically fainter)
objects estimated to be nearer. The object with the smallest estimated
distance is WISEA~J141327.57$+$524831.6 ($\prot \sim 0.8$~d), at \apx90~pc.

The top two panels of \autoref{sample} suggest that our search has fairly
uniform sensitivity across a range of $(i - z)_\text{P1}$ colors, apparent
\zps\ magnitudes, and rotation periods. As the bottom panel of
\autoref{sample} demonstrates, however, several important biases are at work
in the construction of our catalog. Our sample is derived from flux-limited
catalogs and spans a wide range in \teff, inducing color-dependent trends. We
are only sensitive to objects with rotation periods $\gtrsim 0.7$~d and
variability semi-amplitudes $\gtrsim 1$\%, while a search for periodic
variability in the more photometrically stable \textit{Kepler} data set
suggests that the bulk of periodically variable objects have lower amplitudes
\citep{mcquillan2014}, although that data set is dominated by stars hotter
than those in our sample. Finally, we are not able to distinguish multiple
systems from single stars in our catalog.

\autoref{mass-prot} compares our sample to those presented in several
comparable studies \citep{mcquillan2013, mcquillan2014, newton2016}. One
version of the plot differentiates stars based on each study's estimated
stellar masses. However, our masses have \apx20\% uncertainties, and the
masses tabulated by \citet{mcquillan2013, mcquillan2014} are those provided in
the Kepler Input Catalog \citep[KIC;][]{blee11}, which are known to be
systematically biased at low masses \citep{dc13}. We therefore also use
$\text{MEarth} - K_s$ or $\text{pseudoMEarth} - K_s$ color as the abscissa,
where the utility of this quantity is discussed by \citet{dicn16} and the
pseudo-MEarth magnitude is defined by \citet{newton2016}:
\begin{equation}
  \text{pseudoMEarth} = (i_\text{SDSS} + 2z_\text{SDSS}) / 3
    - 0.20\text{ mag}.
\end{equation}
The scatter in the relation is asserted to be \apx5\%. In this version of the
plot we could only include the 211 sources for which we found both SDSS and
2MASS cross-matches. Using the \citet{tonry2012} relations to synthesize SDSS
magnitudes adds another 19 sources but does not change the fundamental
character of the plot.

\autoref{mass-prot} must be interpreted with care because each study is
subject to different selection biases. For instance, it is clear that our
study is less sensitive to fast rotators than the others, which is to be
expected since the other studies are based on data taken more frequently than
our daily cadence. We reiterate that the rotators in our sample with periods
lower than the daily cadence should be considered to be significantly more
uncertain than measurements of longer periods. We are also not sensitive to
low-amplitude rotators. These are responsible for the dense stripes at
relatively long rotation periods in the \textit{Kepler} sample \citep[their
  Figure~4]{mcquillan2014}, which explains why our sample does not show an
analogous feature in \autoref{mass-prot}. We defer a full comparison of the
data sets shown in \autoref{mass-prot} to future work but consider a few
points below.

\subsection{Very Slow Rotators}

In terms of either modeled stellar mass or $\text{(pseudo)MEarth} - K_s$
color, our sample bridges the \textit{Kepler}-based data set of
\citet{mcquillan2013, mcquillan2014}, which generally targeted stars with $M
\gtrsim 0.35$~\msun, and studies of nearby, bright M~dwarfs \citep{irwin2011,
  newton2016} using the MEarth data set, which targeted stars below this
threshold. This is useful because it helps differentiate trends that may be
due to systematic effects from ones that are more likely to be astrophysical.
For instance, \citet{newton2016} noted that the MEarth data set contains a
substantial number of objects with rotation periods $> 70$~d, while
\citet{mcquillan2013, mcquillan2014} found none in \textit{Kepler}. They
suggested that this may lack may have been due to \textit{Kepler} systematics,
especially the effects of the quarterly reorientation of the spacecraft. We
find tentative evidence that this is the case, because our much smaller sample
recovers two objects with $\prot > 70$~d, \object{WISEA~J141100.79$+$541430.7}
and \object{WISEA~J232843.05$+$004453.1}, at $\text{pseudoMEarth} - K_s <
2.2$~mag, a regime in which \textit{Kepler} had good coverage. Analogous
objects should exist in the \textit{Kepler} data set, and the lack of any
detections suggests that searches to date have not been sensitive to them.
With only two such examples, however, firm conclusions cannot yet be drawn.
Based on our data alone, it is possible that only stars with
$\text{(pseudo)MEarth} - K_s \gtrsim 2.0$~mag or $M \lesssim 0.45$~\msun\ are
very slow rotators, but we have very few stars with $\text{(pseudo)MEarth} -
K_s < 2.0$~mag, and our mass estimates are imprecise. It is also possible that
very slowly-rotating stars exist on the blue / high-mass sides of these
thresholds, but that such stars do not possess the nonaxisymmetric
photospheric inhomogeneities that make them discoverable in photometric
rotation period searches.

\subsection{Amplitude of Variability}

\begin{figure}[tb]
  \includegraphics[width=\linewidth]{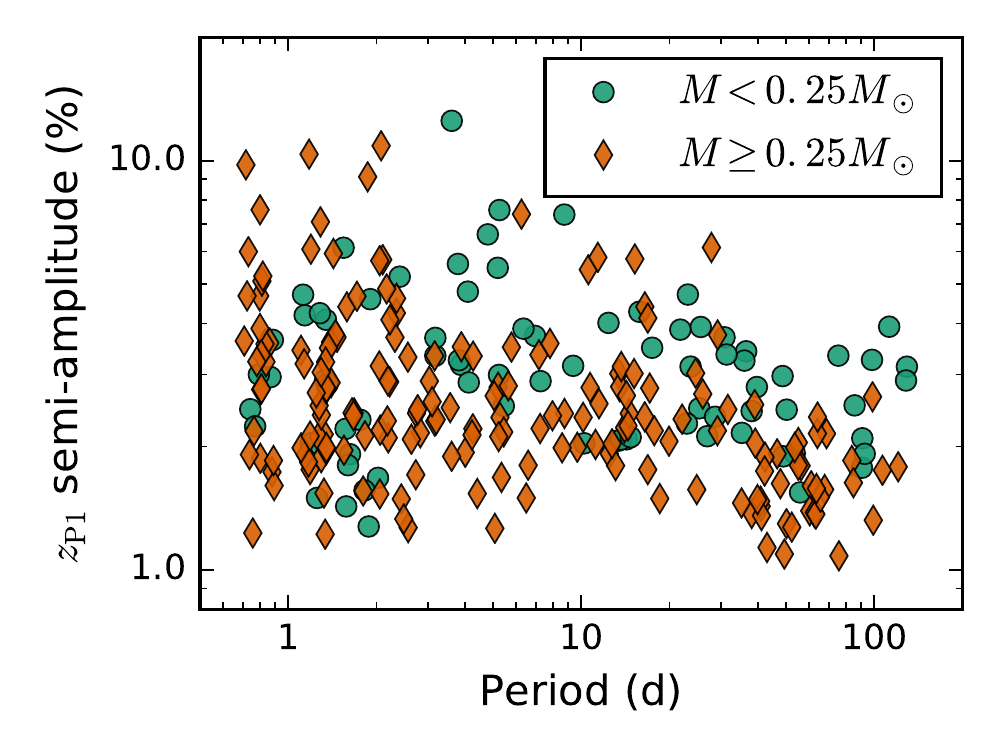}
  \caption{Variability amplitude in the \zps\ band as a function of rotation
    period. The gap in periods at 0.9--1.1~days comes from cuts to avoid false
    detections near the daily observing cadence of PS1-MDS. There is an
    underdensity of low-mass sources at semi-amplitudes $\lesssim 2$\% in the
    period range $3$--$40$~d despite the fact that our search is not unusually
    insensitive to these sources (\autoref{synth_pcolor}).}
\label{protamp}
\end{figure}

\autoref{protamp} shows the relationship between rotation period and
variability semi-amplitude in our final sample. We emulate \citet{newton2016}
by partitioning our targets into those with estimated masses larger or smaller
than 0.25~\msun. While the semi-amplitudes we find are larger than those of
\citet{newton2016}, the shapes of the distributions we find are largely
similar, with a marked lack of high-amplitude slow rotators. A notable
exception is that we observe hints of an anti-correlation between rotation
period and semi-amplitude in both mass bins, while \citet{newton2016} do not
see one in the lower-mass objects. Unlike \citet{newton2016}, however, we have
not isolated a ``statistical sample'' that attempts to account for sensitivity
effects and multiplicity, and high-amplitude and/or short-period objects will
be overrepresented in our sample compared to their true prevalence due to
their relative ease of detection.

\begin{figure}[tb]
  \includegraphics[width=0.9\linewidth]{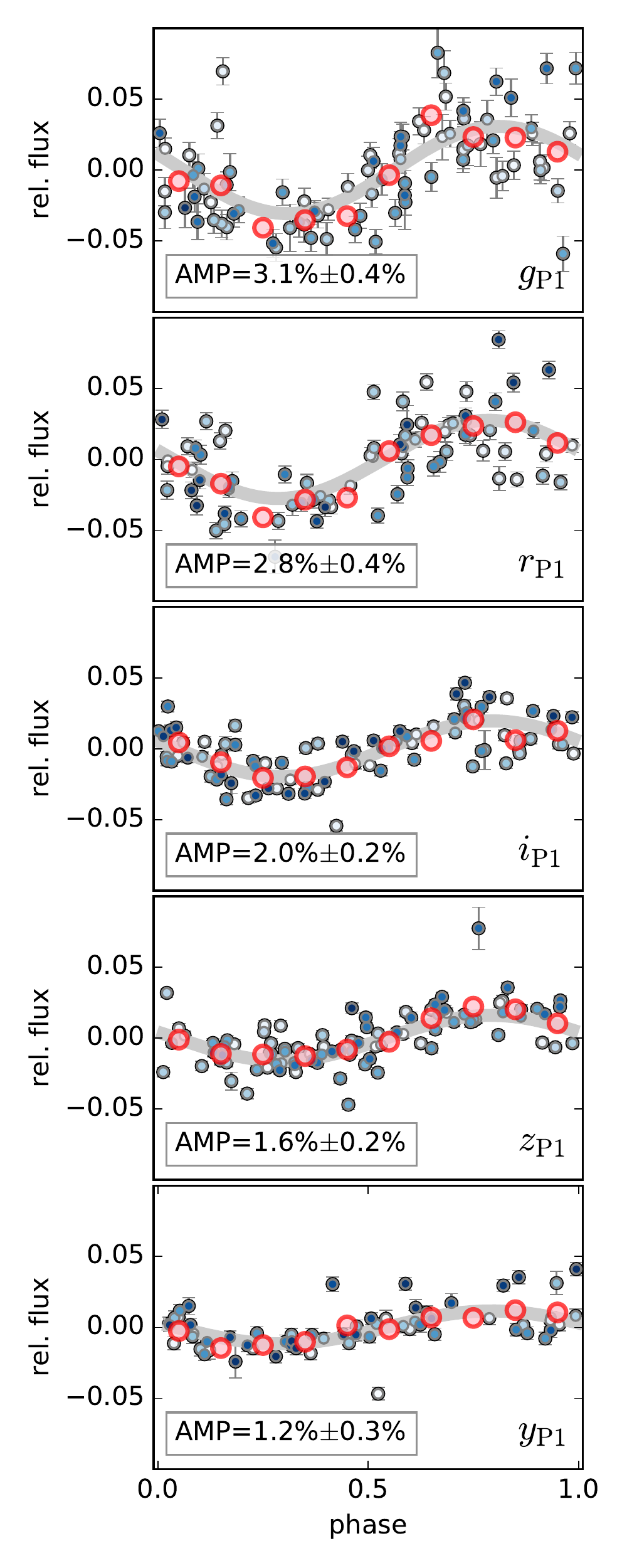}
  \caption{The phased light curve of \object{WISEA~J222234.93$-$001655.4}
    (P$_\text{rot}=1.87$ days) with the amplitude of variability for each
    filter allowed to vary.
    The semi-amplitude of each filter is shown in the bottom left corner
    of each panel.
    The points show the phased light curve of the
    source, colored by time of observation. As in \autoref{samplelightcurves},
    the purple points show the binned mean flux to guide the eye. The model
    for each filter is shown by the gray curve in each panel. The drop in
    amplitude is significant between all filters except between \gps\ to
    \rps. }
  \label{ampvarexample}
\end{figure}

Our sample, like that of \citet{newton2016}, shows a dearth of low-mass
M~dwarfs with intermediate rotation periods and low variability amplitudes.
The larger samples of \citet{mcquillan2013, mcquillan2014} and
\citet{newton2016} have provided compelling evidence that there is a genuine
overall lack of low-mass dwarfs at intermediate rotation periods, reminiscent
of studies of magnetic activity \citep[e.g.,][]{grh02, browning2010} tracing
back to the ``gap'' in \ion{Ca}{2} H\&K emission in F and G stars discovered
by \citet{vaughan1980}. \citet{newton2016} suggest that this gap exists
because low-mass M~dwarfs suddenly undergo rapid spin-down at intermediate
ages. If this hypothesis is correct, this spin-down is apparently associated
with high levels of non-axisymmetric photospheric inhomogeneity. An analogous
gap in rotation periods has been observed in young clusters
\citep[e.g.,][]{mms+11} and is possibly explained by a rapid simplification of
the magnetic field topology that leads to a large increase in angular momentum
loss rates \citep{gdc15}. Such a transition could plausibly lead to large,
stable magnetic spots, consistent with the high-amplitude variability seen
here.

\subsection{Wavelength-Dependence of Variability Amplitude}

Unlike other large surveys for stellar rotational modulation, our study
includes data in five photometric filters, allowing investigation of
wavelength-dependent phenomena in our sample. Inspection of the data shows
that the amplitude of variability for each source can indeed vary
significantly across the five PS1-MDS filters. Because our initial
period-finding analysis assumes a common amplitude across all filters, to
estimate filter-specific amplitudes we re-fit the final sample of rotators
with a simple sinusoidal model, this time allowing the amplitude in each band
to vary independently. \autoref{ampvarexample} shows an example of a light
curve with a significant change in amplitude of variability as a function of
filter. The best-fit rotation period of \object{WISEA~J105130.40$+$572218.9}
is consistent across all filters at 2.31 days. The amplitude of variability,
however, ranges from $1.16\%\pm0.26\%$ in \yps\ to $3.08\%\pm0.41\%$ in \rps.
Across the final sample, 248 out of the 271 stars have significant drops in
amplitude from \rps\ to \yps, with the majority following the trend of
decreasing amplitude in the redder bands. These findings are consistent with
the interpretation that the periodic modulations in our sample of M~dwarf
light curves are caused by starspots with effective temperatures lower than
that of the stellar photosphere \citep{amado2000}. Due to the red colors of
the sources, however, the \textit{absolute} variability amplitudes in the
redder bands tend to be greater than those in the bluer bands.

\subsection{X-ray Activity}

\begin{figure}[tb]
  \includegraphics[width=\linewidth]{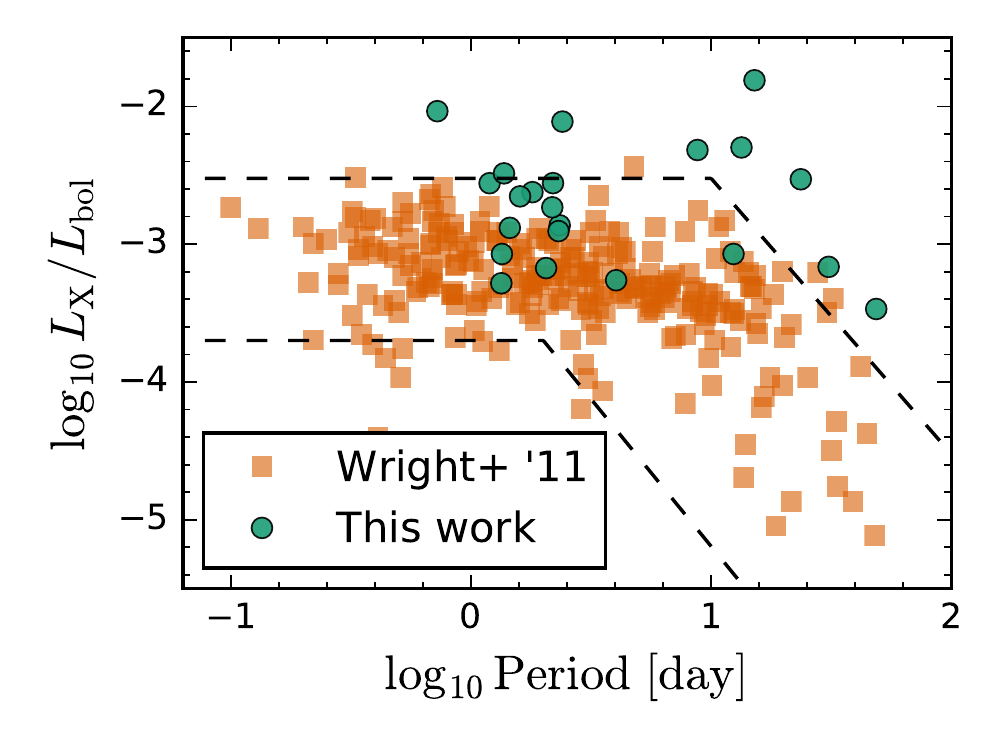}
  \caption{Relation between rotation and X-ray activity for our sample,
    compared to comparable data from \citet{wdmh11}. \textit{Dashed lines}:
    approximate envelope of the \lxlb\ vs. \prot\ relationship observed by
    \citet{wdmh11}.}
  \label{xrot}
\end{figure}

\autoref{xrot} shows the relationship between rotation and X-ray activity in
our sample and compares it to a subset of stars from the compilation of
\citet{wdmh11}. In particular, we show only stars cataloged being field
objects with estimated masses $<$0.6~\msun. The relation between rotation and
X-ray activity past the fully-convective boundary is of particular interest,
and is one where our data set is particularly valuable. While the full catalog
of \citet{wdmh11} includes 824 stars, only 178 of these are field objects with
estimated masses $<$0.35~\msun. Our work contributes 16 such objects.

Rotation/activity analyses often quantify rotation using the Rossby number
$\text{Ro} \equiv \prot / \tau_c$, where $\tau_c$ is nominally a turnover
timescale associated with convective fluid motions in a star. However, in
practice $\tau_c$ is determined by evaluating a function of other stellar
parameters that is tuned to reduce the scatter observed in rotation/activity
relations, so its use injects an ill-defined model dependence into subsequent
analysis \citep[cf.][]{rsp14}. We therefore simply quantify rotation with
\prot. If we reproduce \autoref{xrot} using Ro to quantify rotation, no trends
in the data emerge.

Quantifying X-ray activity with \lxlb, rather than $L_\text{X}$, injects an
additional model dependence into our analysis, since $\lxlb =
f_\text{X}/f_\text{bol}$ depends on our computation of the \zps-band
bolometric correction factor $\text{BC}_{z,\text{P1}}$. However, $L_\text{X}$
is a less ideal choice in this case since computing it requires distance
measurements, which in our sample are less accurate estimates based on
color-magnitude relations.

\input{xray}

\autoref{xrot} is notable for showing no clear relation between rotation and
activity in our sample, even though our data span nearly two orders of
magnitude in rotation period. However, our sample is strongly biased because
it contains no X-ray nondetections by construction. Given the
characteristically large distances of our targets, only the X-ray brightest of
them will have been detected in our catalog search. The source with the
largest observed \lxlb, \object{WISEA~J161112.96$+$541508.3} ($\prot =
15.1$~d), does not otherwise appear unusual. While it is not flagged as an
X-ray variable in its 3XMM-DR5 catalog record, it was detected in just two
separate XMM observations separated by \apx40~days and so could plausibly have
been observed during flares, implying a true quiescent flux lower than what is
cataloged. The slowest-rotating target with an X-ray detection,
\object{WISEA~J160956.04$+$543646.9} ($\prot = 48.7$~d), is a weak detection
with $f_X = (8 \pm 4) \times 10^{-16}$~erg~s$^{-1}$~cm$^{-2}$. Its estimated
Rossby number is $\approx$0.6, placing it in the ``unsaturated'' regime of the
standard rotation/activity relation \citep{wdmh11} but not in a location that
makes its X-ray brightness unusual.

\section{Conclusions} \label{conclusions}

We have measured the rotation periods of 271 cool stars in the PS1-MDS
footprint, making a substantial contribution to the overall sample of
fully-convective dwarfs with measured photometric rotation periods.

The PS1-MDS data set has strengths that differentiate it from the other data
sets used to perform comparable studies. Because our photometry spans five
filters, we can winnow a large (\apx4~million source) input catalog to a
tractable list of sources of interest, as well as study the
wavelength-dependence of the photospheric structures that lead to periodic
variability and search for objects that are unusual in this regard. Moreover,
because the MDS pointings overlap various fields with deep multiwavelength
coverage (GOODS-South, the Lockman Hole, etc.), large amounts of potentially
useful archival data are publicly available. We demonstrated this approach via
a cross-match of our catalog to several well-known X-ray catalogs, finding 22
matches; the effort here, however, just scratches the surface of what is
possible. In the more distant future, comparable analysis of the LSST data
stream \citep{the.lsst} will yield a sample of stellar rotators orders of
magnitude larger than the one presented here, allowing true statistical
analysis of the relationship between rotation and other parameters in cool
field dwarfs.

To enable basic comparisons to prior work, we have extended the technique of
\citet{mann2015, mfg+16.erratum} to develop new polynomials relating PS1
colors to stellar properties and applied these relations to our targets. Now
that these stars have been identified as being of unusual interest, it would
be valuable to characterize them more precisely and accurately. Both archival
investigations and targeted follow-up are underway.

In this work we have focused on developing a catalog of sources that can be
confidently identified as cool dwarfs with periodic photometric variability.
We have \textit{not} attempted to characterize the sensitivity of our search
in detail, nor have we attempted to search for potentially-interesting but
marginal candidates. The foundation of such detailed characterization will be
a robust, well-understood photometric extraction pipeline, and we therefore
defer such work until the final ``PV3'' reprocessing of the MDS nightly stacks
has been performed, at which point we will re-run our search. Final
classification of variables will be fully automated to allow us to
characterize the search's sensitivity through simulations.

Although neither our search sensitivity nor the parameters of our targets are
fully characterized, the catalog presented in this work shows several
interesting features. First, we find tentative evidence that \textit{Kepler}
searches for photometric rotation periods are systematically insensitive to
objects with $\prot > 70$~d, as suggested by \citet{newton2016}. Second, we
note a dearth of low-mass ($< 0.25$~\msun) objects with intermediate
(\apx10--40~d) rotation periods and low ($< 2$\%) variability semi-amplitudes.
If the overall lack of such objects is indeed due to a rapid evolution at
intermediate \citep[2--5~Gyr;][]{newton2016} ages, this suggests that such
transitional objects have relatively large, nonaxisymmetric photospheric
structures. This may be consistent with the emergence of a less-complex
magnetic topology, as proposed by \citet{gdc15} to explain a similar ``period
gap'' in young stars.

\acknowledgments

\textit{Acknowledgments.} We thank Elisabeth Newton for her insight into the
MEarth dataset. This work was supported in part by the National Science
Foundation REU and Department of Defense ASSURE programs under NSF Grant no.
1262851 and by the Smithsonian Institution.

The Pan-STARRS1 Surveys (PS1) have been made possible through contributions of
the Institute for Astronomy, the University of Hawaii, the Pan-STARRS Project
Office, the Max-Planck Society and its participating institutes, the Max
Planck Institute for Astronomy, Heidelberg and the Max Planck Institute for
Extraterrestrial Physics, Garching, The Johns Hopkins University, Durham
University, the University of Edinburgh, Queen's University Belfast, the
Harvard-Smithsonian Center for Astrophysics, the Las Cumbres Observatory
Global Telescope Network Incorporated, the National Central University of
Taiwan, the Space Telescope Science Institute, the National Aeronautics and
Space Administration under Grant No. NNX08AR22G issued through the Planetary
Science Division of the NASA Science Mission Directorate, the National Science
Foundation under Grant No. AST-1238877, the University of Maryland, and Eotvos
Lorand University (ELTE) and the Los Alamos National Laboratory.

\bibliography{\jobname}

\appendix

\section{Estimating M~Dwarf Parameters from PS1 Photometry}
\label{stellar_properties}

Here we provide a more detailed description of our method for determining
M~dwarf stellar parameters from their PS1 photometry
(\autoref{stellar_properties_summary}).

\citet{mann2015} provide relations between SDSS $griz$ photometry and
fundamental properties of M dwarfs (e.g., \teff, $R_*$) based on a sample of
183 nearby M dwarfs with precise parallaxes and flux-calibrated optical and
NIR spectra. It is possible to convert these to relations to PS1 photometry
using the transformations in \citet{tonry2012}. However, the majority of the
sample in this paper has $(r{-}i)_\text{P1} > 1$, while only a small fraction
of the SEDs used to derive the SDSS--PS1 transformations in \citet{tonry2012}
are this red. Instead we derive new relations using synthetic PS1 magnitudes.
We convolve the filter profiles from \citet{tonry2012} with the
\citet{mann2015} spectra and convert the resulting fluxes to PS1 magnitudes
using the appropriate zero points and ``tweaks'' (used to force agreement
between PS1 photometry and spectroscopic standards) from \citet{tonry2012}. We
then derive new polynomial relations between \teff, BC$_{z,\text{P1}}$,
$M_{z,\text{P1}}$, and $M_*$ and our synthetic $(g{-}i)_\text{P1}$,
$(r{-}i)_\text{P1}$, $(i{-}z)_\text{P1}$, or $(r{-}y)_\text{P1}$ colors.
Additional terms are added to the polynomial relations as long as they are
justified by an F-test. To help account for systematics due to metallicity we
also derive relations that include $(g{-}r)_\text{P1}$ in addition to one of
the above colors for all physical properties except $BC_{z,\text{P1}}$, which
shows negligible improvement from the additional term. We note that these
color combinations are imperfect measures of [Fe/H] and the relations likely
still have systematics with metallicity.

% The difference between synthetic PS1 and SDSS magnitudes derived from the
% same spectra diverge from those given in \citet{tonry2012}, particularly for
% the $z$ and $y$ bands, but are in agreement with the offsets seen in the
% sample of stars in this paper with both SDSS and PS1 photometry.

\input{mann}

In \autoref{t.mann} we present coefficients for the new polynomials we have
derived, where each row corresponds to an equation of the form
\begin{equation}
  \frac{y}{\text{Unit}} = \sum_{i \geq 0} a_i \left(\frac{C_a}{\text{mag}}\right)^i +
    \sum_{i \geq 1} b_i \left(\frac{C_b}{\text{mag}}\right)^i,
  \label{e.mann_polynomial}
\end{equation}
and the second term is only included for rows that list the $C_b$ parameter.
Note that we always have $b_0 = 0$. The rms scatters of the various polynomial
fits to the properties of the \citet{mann2015} sample are tabulated as well.

To derive parameters for our stars, we used the polynomials that are functions
of $(i{-}z)_\text{P1}$, using the metallicity-correcting variants that also
depend on $(g{-}r)_\text{P1}$ for all of the objects in our sample except the
two that do not have a measured $(g{-}r)_\text{P1}$ color.

Before evaluating these polynomials, we first iteratively estimated each
target star's distance and reddening in the following manner. We derived an
initial distance estimate from each star's reddened PS1 colors by computing
its distance modulus $\text{DM} \equiv 5 \log_{10} (d / 10\text{ pc})$ from
the observed data and our polynomial expression for $M_{z,\text{P1}}$. We then
estimated its reddening from the 3D dust maps of \citet{gsf+15}, using the
``Argonaut'' web service provided by \citet{gsf+15} to download the estimated
3D dust distribution for each relevant line of sight. This estimate is
expressed as a set of 20 curves sampling the growth of the reddening parameter
$E(B-V)$ as a function of distance modulus, each curve representing one sample
from the \citet{gsf+15} MCMC analysis of the colors and estimated distances of
binned groups of stars. For each of the 20 curves we generated 50 samples of
$E(B-V)$ by perturbing the DM estimate by a random offset drawn from a
Gaussian distribution corresponding to the noise in the $M_{z,\text{P1}}$
polynomial relation. In this analysis we used piecewise linear interpolation
between the curve sample points because the shape of the $E(B-V)$ growth
function was often jagged, leading to poor fits when cubic spline
interpolation was used.

Given a value of $E(B-V)$, the reddening in the PS1 filters can be determined
by multiplying by one of the constants tabulated by \citet{sf11}. For each
source of interest, we iteratively computed reddening corrections and distance
estimates until the fractional difference between distance moduli estimated in
subsequent iterations was less than 0.001. This process converged for all
sources for which it was attempted, although in some cases the estimated
distance modulus was not in the range indicated as ``reliable'' in the
\citet{gsf+15} data set. We used the scatter in the 1000 samples of $E(B-V)$
to determine uncertainties in the derived distances and reddening corrections.
In the final catalog the median $A_V$ is 0.04~mag and the 95th percentile is
0.22~mag, and typical distance uncertainties are 20\%. While, as discussed in
\autoref{contaminants}, our procedure will result in ${\lesssim}\sqrt{2}$
errors for spatially unresolved binaries, these errors are comparable to the
overall accuracy of our procedure given the available information.

To compute \teff, $M$, and $f_\text{bol}$, we evaluated the polynomials given
above using the dereddened magnitudes. The apparent bolometric magnitudes are
converted to fluxes assuming an absolute solar bolometric magnitude of $4.7554
\pm 0.0004$~mag and solar luminosity of $(3.8270 \pm 0.0014) \times
10^{33}$~erg~s$^{-1}$ \citep{m12, pm13}.

\end{document}

%% file: rotators-sample.tex
% TableBuilder table
\begin{deluxetable}{lr@{}l@{\,}lr@{}l@{\,}lr@{}l@{\,}lr@{}l@{\,}lr@{}l@{\,}lr@{}l@{\,}lr@{}l@{\,}lr@{}l@{\,}lr@{}l@{\,}lcr@{}l@{\,}lr@{}lr@{}lc}
%custom preamble
\tabletypesize{\tiny}\rotate
%hardcoded preamble
\tablecolumns{37}
\tablewidth{0em}
\tablecaption{Candidate Rotating M~Dwarfs\tablenotemark{a}\label{t.rotators}}
\tablehead{
\colhead{Name} & \multicolumn{3}{c}{$z_\text{P1}$} & \multicolumn{3}{c}{$(g-r)_\text{P1}$} & \multicolumn{3}{c}{$(r-i)_\text{P1}$} & \multicolumn{3}{c}{$(i-z)_\text{P1}$} & \multicolumn{3}{c}{$(z-y)_\text{P1}$} & \multicolumn{3}{c}{\teff} & \multicolumn{3}{c}{Mass} & \multicolumn{3}{c}{$d$} & \multicolumn{3}{c}{$\log_{10} L_\text{bol}$} & \colhead{Giant?} & \multicolumn{3}{c}{$P_\text{rot}$} & \multicolumn{2}{c}{Ampl.} & \multicolumn{2}{c}{Pk. Rat.} & \colhead{\paeight} \\
 & \multicolumn{3}{c}{(mag)} & \multicolumn{3}{c}{(mag)} & \multicolumn{3}{c}{(mag)} & \multicolumn{3}{c}{(mag)} & \multicolumn{3}{c}{(mag)} & \multicolumn{3}{c}{(K)} & \multicolumn{3}{c}{(M$_\odot$)} & \multicolumn{3}{c}{(pc)} & \multicolumn{3}{c}{[$L_\odot$]} &  & \multicolumn{3}{c}{(day)} & \multicolumn{2}{c}{(\%)} &  &  &  \\ \\
\multicolumn{1}{c}{(1)} & \multicolumn{3}{c}{(2)} & \multicolumn{3}{c}{(3)} & \multicolumn{3}{c}{(4)} & \multicolumn{3}{c}{(5)} & \multicolumn{3}{c}{(6)} & \multicolumn{3}{c}{(7)} & \multicolumn{3}{c}{(8)} & \multicolumn{3}{c}{(9)} & \multicolumn{3}{c}{(10)} & \multicolumn{1}{c}{(11)} & \multicolumn{3}{c}{(12)} & \multicolumn{2}{c}{(13)} & \multicolumn{2}{c}{(14)} & \multicolumn{1}{c}{(15)}
}
\startdata
WISEA J021823.69$-$044932.5 & $17$ & $.72$ & $\pm\,0.03$\rule{0pt}{3ex} & $1$ & $.29$ & $\pm\,0.05$\rule{0pt}{3ex} & $1$ & $.65$ & $\pm\,0.05$\rule{0pt}{3ex} & $0$ & $.75$ & $\pm\,0.05$\rule{0pt}{3ex} & $0$ & $.39$ & $\pm\,0.05$\rule{0pt}{3ex} & $3130$ &  & $\pm\,80$\rule{0pt}{3ex} & $0$ & $.21$ & $\pm\,0.04$\rule{0pt}{3ex} & $350$ &  & $\pm\,70$\rule{0pt}{3ex} & $-2$ & $.3$ & $\pm\,0.2$\rule{0pt}{3ex} &  & $13$ & $.37$ & $\pm\,0.03$\rule{0pt}{3ex} & $2$ & $.1$ & $1$ & $.71$ & 12 \\
WISEA J021907.51$-$033114.2 & $16$ & $.74$ & $\pm\,0.03$\rule{0pt}{3ex} & $1$ & $.32$ & $\pm\,0.05$\rule{0pt}{3ex} & $1$ & $.77$ & $\pm\,0.05$\rule{0pt}{3ex} & $0$ & $.80$ & $\pm\,0.05$\rule{0pt}{3ex} & $0$ & $.41$ & $\pm\,0.05$\rule{0pt}{3ex} & $3060$ &  & $\pm\,80$\rule{0pt}{3ex} & $0$ & $.17$ & $\pm\,0.03$\rule{0pt}{3ex} & $190$ &  & $\pm\,40$\rule{0pt}{3ex} & $-2$ & $.4$ & $\pm\,0.2$\rule{0pt}{3ex} &  & $12$ & $.39$ & $\pm\,0.03$\rule{0pt}{3ex} & $4$ & $.0$ & $2$ & $.30$ & 38 \\
WISEA J022042.16$-$030701.0 & $16$ & $.38$ & $\pm\,0.03$\rule{0pt}{3ex} & $1$ & $.22$ & $\pm\,0.05$\rule{0pt}{3ex} & $1$ & $.10$ & $\pm\,0.05$\rule{0pt}{3ex} & $0$ & $.52$ & $\pm\,0.05$\rule{0pt}{3ex} & $0$ & $.23$ & $\pm\,0.05$\rule{0pt}{3ex} & $3470$ &  & $\pm\,80$\rule{0pt}{3ex} & $0$ & $.40$ & $\pm\,0.07$\rule{0pt}{3ex} & $390$ &  & $\pm\,80$\rule{0pt}{3ex} & $-1$ & $.7$ & $\pm\,0.2$\rule{0pt}{3ex} &  & $2$ & $.195$ & $\pm\,0.003$\rule{0pt}{3ex} & $2$ & $.1$ & $1$ & $.37$ & 26 \\
WISEA J022049.11$-$041237.1 & $19$ & $.41$ & $\pm\,0.03$\rule{0pt}{3ex} & $1$ & $.20$ & $\pm\,0.06$\rule{0pt}{3ex} & $1$ & $.66$ & $\pm\,0.05$\rule{0pt}{3ex} & $0$ & $.76$ & $\pm\,0.05$\rule{0pt}{3ex} & $0$ & $.38$ & $\pm\,0.05$\rule{0pt}{3ex} & $3160$ &  & $\pm\,80$\rule{0pt}{3ex} & $0$ & $.24$ & $\pm\,0.04$\rule{0pt}{3ex} & $740$ &  & $\pm\,100$\rule{0pt}{3ex} & $-2$ & $.3$ & $\pm\,0.2$\rule{0pt}{3ex} & ? & $25$ & $.57$ & $\pm\,0.10$\rule{0pt}{3ex} & $3$ & $.9$ & $1$ & $.84$ & 22 \\
WISEA J022106.66$-$033528.4 & $17$ & $.72$ & $\pm\,0.03$\rule{0pt}{3ex} & $1$ & $.27$ & $\pm\,0.05$\rule{0pt}{3ex} & $1$ & $.45$ & $\pm\,0.05$\rule{0pt}{3ex} & $0$ & $.65$ & $\pm\,0.05$\rule{0pt}{3ex} & $0$ & $.30$ & $\pm\,0.05$\rule{0pt}{3ex} & $3260$ &  & $\pm\,80$\rule{0pt}{3ex} & $0$ & $.28$ & $\pm\,0.05$\rule{0pt}{3ex} & $480$ &  & $\pm\,90$\rule{0pt}{3ex} & $-2$ & $.0$ & $\pm\,0.2$\rule{0pt}{3ex} & ? & $16$ & $.48$ & $\pm\,0.05$\rule{0pt}{3ex} & $4$ & $.4$ & $1$ & $.91$ & 41 \\
WISEA J022113.86$-$052801.8 & $18$ & $.31$ & $\pm\,0.03$\rule{0pt}{3ex} & $1$ & $.34$ & $\pm\,0.05$\rule{0pt}{3ex} & $1$ & $.76$ & $\pm\,0.05$\rule{0pt}{3ex} & $0$ & $.83$ & $\pm\,0.05$\rule{0pt}{3ex} & $0$ & $.37$ & $\pm\,0.05$\rule{0pt}{3ex} & $3030$ &  & $\pm\,80$\rule{0pt}{3ex} & $0$ & $.16$ & $\pm\,0.03$\rule{0pt}{3ex} & $360$ &  & $\pm\,70$\rule{0pt}{3ex} & $-2$ & $.5$ & $\pm\,0.2$\rule{0pt}{3ex} & ? & $1$ & $.6248$ & $\pm\,0.0006$\rule{0pt}{3ex} & $1$ & $.9$ & $2$ & $.91$ & 12 \\
WISEA J022121.80$-$043127.9 & $16$ & $.58$ & $\pm\,0.03$\rule{0pt}{3ex} & $1$ & $.27$ & $\pm\,0.05$\rule{0pt}{3ex} & $1$ & $.44$ & $\pm\,0.05$\rule{0pt}{3ex} & $0$ & $.66$ & $\pm\,0.05$\rule{0pt}{3ex} & $0$ & $.30$ & $\pm\,0.04$\rule{0pt}{3ex} & $3240$ &  & $\pm\,80$\rule{0pt}{3ex} & $0$ & $.27$ & $\pm\,0.05$\rule{0pt}{3ex} & $280$ &  & $\pm\,50$\rule{0pt}{3ex} & $-2$ & $.0$ & $\pm\,0.2$\rule{0pt}{3ex} &  & $1$ & $.3370$ & $\pm\,0.0003$\rule{0pt}{3ex} & $1$ & $.2$ & $1$ & $.23$ & 20 \\
WISEA J022253.37$-$043254.3 & $16$ & $.87$ & $\pm\,0.03$\rule{0pt}{3ex} & $1$ & $.21$ & $\pm\,0.05$\rule{0pt}{3ex} & $0$ & $.84$ & $\pm\,0.05$\rule{0pt}{3ex} & $0$ & $.39$ & $\pm\,0.05$\rule{0pt}{3ex} & $0$ & $.19$ & $\pm\,0.05$\rule{0pt}{3ex} & $3700$ &  & $\pm\,80$\rule{0pt}{3ex} & $0$ & $.53$ & $\pm\,0.10$\rule{0pt}{3ex} & $720$ &  & $\pm\,100$\rule{0pt}{3ex} & $-1$ & $.3$ & $\pm\,0.2$\rule{0pt}{3ex} &  & $5$ & $.203$ & $\pm\,0.004$\rule{0pt}{3ex} & $2$ & $.8$ & $5$ & $.59$ & 25 \\
WISEA J022325.43$-$052529.4 & $17$ & $.03$ & $\pm\,0.03$\rule{0pt}{3ex} & $1$ & $.07$ & $\pm\,0.05$\rule{0pt}{3ex} & $0$ & $.68$ & $\pm\,0.05$\rule{0pt}{3ex} & $0$ & $.34$ & $\pm\,0.05$\rule{0pt}{3ex} & $0$ & $.20$ & $\pm\,0.05$\rule{0pt}{3ex} & $3890$ &  & $\pm\,80$\rule{0pt}{3ex} & $0$ & $.69$ & $\pm\,0.10$\rule{0pt}{3ex} & $900$ &  & $\pm\,200$\rule{0pt}{3ex} & $-1$ & $.2$ & $\pm\,0.2$\rule{0pt}{3ex} & P & $36$ & $.2$ & $\pm\,0.3$\rule{0pt}{3ex} & $5$ & $.2$ & $1$ & $.93$ & 19 \\
WISEA J022327.16$-$052055.0 & $15$ & $.85$ & $\pm\,0.03$\rule{0pt}{3ex} & $1$ & $.22$ & $\pm\,0.05$\rule{0pt}{3ex} & $1$ & $.42$ & $\pm\,0.05$\rule{0pt}{3ex} & $0$ & $.65$ & $\pm\,0.05$\rule{0pt}{3ex} & $0$ & $.30$ & $\pm\,0.05$\rule{0pt}{3ex} & $3290$ &  & $\pm\,80$\rule{0pt}{3ex} & $0$ & $.30$ & $\pm\,0.05$\rule{0pt}{3ex} & $200$ &  & $\pm\,40$\rule{0pt}{3ex} & $-2$ & $.0$ & $\pm\,0.2$\rule{0pt}{3ex} &  & $4$ & $.019$ & $\pm\,0.003$\rule{0pt}{3ex} & $1$ & $.9$ & $4$ & $.10$ & 12 \\
WISEA J022400.58$-$052005.1 & $16$ & $.36$ & $\pm\,0.03$\rule{0pt}{3ex} & $1$ & $.15$ & $\pm\,0.05$\rule{0pt}{3ex} & $1$ & $.40$ & $\pm\,0.05$\rule{0pt}{3ex} & $0$ & $.62$ & $\pm\,0.05$\rule{0pt}{3ex} & $0$ & $.31$ & $\pm\,0.05$\rule{0pt}{3ex} & $3370$ &  & $\pm\,80$\rule{0pt}{3ex} & $0$ & $.37$ & $\pm\,0.07$\rule{0pt}{3ex} & $280$ &  & $\pm\,60$\rule{0pt}{3ex} & $-1$ & $.9$ & $\pm\,0.2$\rule{0pt}{3ex} &  & $60$ & $.9$ & $\pm\,0.5$\rule{0pt}{3ex} & $1$ & $.6$ & $1$ & $.08$ & 11 \\
WISEA J022411.77$-$041527.7 & $17$ & $.86$ & $\pm\,0.03$\rule{0pt}{3ex} & \multicolumn{3}{c}{}\rule{0pt}{3ex} & \multicolumn{3}{c}{}\rule{0pt}{3ex} & $0$ & $.74$ & $\pm\,0.05$\rule{0pt}{3ex} & $0$ & $.27$ & $\pm\,0.05$\rule{0pt}{3ex} & $3170$ &  & $\pm\,90$\rule{0pt}{3ex} & $0$ & $.23$ & $\pm\,0.04$\rule{0pt}{3ex} & $380$ &  & $\pm\,80$\rule{0pt}{3ex} & $-2$ & $.3$ & $\pm\,0.2$\rule{0pt}{3ex} & ? & $1$ & $.3451$ & $\pm\,0.0010$\rule{0pt}{3ex} & $4$ & $.1$ & $1$ & $.22$ & 28 \\
\enddata
\tablecomments{Missing values in columns (3) through (6) may occur if the deep photometric
extraction failed due to missing data or an inability to determine the
absolute photometric calibration. Columns (7) through (10) are estimated from
colors as described in \autoref{stellar_properties}. A ``P'' in column (11)
indicates a probable giant; a ``?'' indicates a source missing the robust
2MASS and/or WISE photometry needed to assess the criterion. Column (13) is
the semi-amplitude of the periodic variability signal. Columns (14) and (15)
are the periodogram significance metrics discussed in
\autoref{periodsearch}.}
\tablenotetext{a}{This is a sample version of this table showing only the first few rows.
    The full table is available in machine-readable form.}
\end{deluxetable}
% end TableBuilder table

%% file: xray.tex
% TableBuilder table
\begin{deluxetable*}{llr@{}l@{\,}lr@{}l@{\,}lrr@{}l@{\,}lr@{}l@{\,}l}
%custom preamble

%hardcoded preamble
\tablecolumns{15}
\tablewidth{0em}
\tablecaption{Rotators with Archival X-Ray Detections\label{t.xray}}
\tablehead{
\colhead{Name} & \colhead{X-Ray Name} & \multicolumn{3}{c}{Mass} & \multicolumn{3}{c}{$P_\text{rot}$} & \colhead{$\Delta\theta$} & \multicolumn{3}{c}{$f_X$} & \multicolumn{3}{c}{$\log_{10} L_X/L_\text{bol}$} \\
 &  & \multicolumn{3}{c}{(M$_\odot$)} & \multicolumn{3}{c}{(day)} & \colhead{(arcsec)} & \multicolumn{3}{c}{(erg s$^{-1}$ cm$^{-2}$)} & \multicolumn{3}{c}{(dex)} \\ \\
\multicolumn{1}{c}{(1)} & \multicolumn{1}{c}{(2)} & \multicolumn{3}{c}{(3)} & \multicolumn{3}{c}{(4)} & \multicolumn{1}{c}{(5)} & \multicolumn{3}{c}{(6)} & \multicolumn{3}{c}{(7)}
}
\startdata
WISEA J021823.69$-$044932.5 & 3XMM J021823.6$-$044931 & $0$ & $.21$ & $\pm\,0.04$\rule{0pt}{3ex} & $13$ & $.37$ & $\pm\,0.03$\rule{0pt}{3ex} & 1.2 & $($$7$ &  & $\pm\,3$$) \times 10^{-15}$\rule{0pt}{3ex} & $-2$ & $.4$ & $\pm\,0.3$\rule{0pt}{3ex}\\ % \\
WISEA J021907.51$-$033114.2 & 2XLSSd J021907.5$-$033114 & $0$ & $.17$ & $\pm\,0.03$\rule{0pt}{3ex} & $12$ & $.38$ & $\pm\,0.03$\rule{0pt}{3ex} & 0.6 & $($$3$ & $.2$ & $\pm\,1.0$$) \times 10^{-15}$\rule{0pt}{3ex} & $-3$ & $.1$ & $\pm\,0.2$\rule{0pt}{3ex}\\ % \\
WISEA J022042.16$-$030701.0 & 3XMM J022042.0$-$030705 & $0$ & $.40$ & $\pm\,0.07$\rule{0pt}{3ex} & $2$ & $.195$ & $\pm\,0.003$\rule{0pt}{3ex} & 5.1 & $($$1$ & $.4$ & $\pm\,0.9$$) \times 10^{-14}$\rule{0pt}{3ex} & $-2$ & $.7$ & $\pm\,0.5$\rule{0pt}{3ex}\\ % \\
WISEA J022121.80$-$043127.9 & 2XLSSd J022121.7$-$043129 & $0$ & $.27$ & $\pm\,0.05$\rule{0pt}{3ex} & $1$ & $.3370$ & $\pm\,0.0003$\rule{0pt}{3ex} & 1.7 & $($$2$ & $.1$ & $\pm\,0.6$$) \times 10^{-15}$\rule{0pt}{3ex} & $-3$ & $.30$ & $\pm\,0.10$\rule{0pt}{3ex}\\ % \\
WISEA J022327.16$-$052055.0 & 2XLSSd J022326.9$-$052056 & $0$ & $.30$ & $\pm\,0.05$\rule{0pt}{3ex} & $4$ & $.019$ & $\pm\,0.003$\rule{0pt}{3ex} & 3.2 & $($$4$ & $.3$ & $\pm\,1.0$$) \times 10^{-15}$\rule{0pt}{3ex} & $-3$ & $.3$ & $\pm\,0.2$\rule{0pt}{3ex}\\ % \\
WISEA J022411.77$-$041527.7 & 3XMM J022411.8$-$041527 & $0$ & $.23$ & $\pm\,0.04$\rule{0pt}{3ex} & $1$ & $.3451$ & $\pm\,0.0010$\rule{0pt}{3ex} & 1.8 & $($$1$ & $.3$ & $\pm\,0.8$$) \times 10^{-15}$\rule{0pt}{3ex} & $-3$ & $.1$ & $\pm\,0.4$\rule{0pt}{3ex}\\ % \\
WISEA J022511.68$-$050503.2 & 3XMM J022511.6$-$050503 & $0$ & $.25$ & $\pm\,0.04$\rule{0pt}{3ex} & $2$ & $.054$ & $\pm\,0.002$\rule{0pt}{3ex} & 0.4 & $($$4$ &  & $\pm\,2$$) \times 10^{-15}$\rule{0pt}{3ex} & $-3$ & $.2$ & $\pm\,0.3$\rule{0pt}{3ex}\\ % \\
WISEA J084921.27$+$444949.2 & CXO J084921.2$+$444949 & $0$ & $.25$ & $\pm\,0.05$\rule{0pt}{3ex} & $2$ & $.3399$ & $\pm\,0.0010$\rule{0pt}{3ex} & 0.2 & $($$3$ & $.7$ & $\pm\,0.7$$) \times 10^{-15}$\rule{0pt}{3ex} & $-2$ & $.87$ & $\pm\,0.08$\rule{0pt}{3ex}\\ % \\
WISEA J095900.97$+$020830.5 & CXO J095900.9$+$020830 & $0$ & $.48$ & $\pm\,0.09$\rule{0pt}{3ex} & $2$ & $.1793$ & $\pm\,0.0008$\rule{0pt}{3ex} & 0.1 & $($$6$ & $.1$ & $\pm\,0.9$$) \times 10^{-15}$\rule{0pt}{3ex} & $-2$ & $.74$ & $\pm\,0.06$\rule{0pt}{3ex}\\ % \\
WISEA J095918.34$+$024304.8 & CXO J095918.3$+$024305 & $0$ & $.65$ & $\pm\,0.10$\rule{0pt}{3ex} & $1$ & $.4502$ & $\pm\,0.0004$\rule{0pt}{3ex} & 0.5 & $($$1$ & $.0$ & $\pm\,0.3$$) \times 10^{-14}$\rule{0pt}{3ex} & $-2$ & $.89$ & $\pm\,0.10$\rule{0pt}{3ex}\\ % \\
WISEA J100052.90$+$015714.1 & CXO J100052.9$+$015714 & $0$ & $.10$ & $\pm\,0.02$\rule{0pt}{3ex} & $23$ & $.61$ & $\pm\,0.10$\rule{0pt}{3ex} & 0.4 & $($$5$ & $.3$ & $\pm\,0.8$$) \times 10^{-15}$\rule{0pt}{3ex} & $-2$ & $.53$ & $\pm\,0.07$\rule{0pt}{3ex}\\ % \\
WISEA J104541.81$+$592041.1 & CXO J104541.9$+$592040 & $0$ & $.20$ & $\pm\,0.04$\rule{0pt}{3ex} & $30$ & $.8$ & $\pm\,0.2$\rule{0pt}{3ex} & 1.1 & $($$4$ & $.2$ & $\pm\,1.0$$) \times 10^{-15}$\rule{0pt}{3ex} & $-3$ & $.2$ & $\pm\,0.2$\rule{0pt}{3ex}\\ % \\
WISEA J104633.88$+$574103.6 & 3XMM J104634.0$+$574103 & $0$ & $.32$ & $\pm\,0.06$\rule{0pt}{3ex} & $8$ & $.77$ & $\pm\,0.02$\rule{0pt}{3ex} & 1.2 & $($$4$ &  & $\pm\,2$$) \times 10^{-14}$\rule{0pt}{3ex} & $-2$ & $.4$ & $\pm\,0.4$\rule{0pt}{3ex}\\ % \\
WISEA J104946.22$+$573026.7 & 3XMM J104946.1$+$573030 & $0$ & $.27$ & $\pm\,0.05$\rule{0pt}{3ex} & $0$ & $.7239$ & $\pm\,0.0002$\rule{0pt}{3ex} & 3.3 & $($$1$ & $.6$ & $\pm\,0.8$$) \times 10^{-14}$\rule{0pt}{3ex} & $-2$ & $.1$ & $\pm\,0.4$\rule{0pt}{3ex}\\ % \\
WISEA J105130.40$+$572218.9 & 3XMM J105130.3$+$572219 & $0$ & $.44$ & $\pm\,0.08$\rule{0pt}{3ex} & $2$ & $.313$ & $\pm\,0.002$\rule{0pt}{3ex} & 0.3 & $($$4$ & $.6$ & $\pm\,1.0$$) \times 10^{-15}$\rule{0pt}{3ex} & $-2$ & $.93$ & $\pm\,0.10$\rule{0pt}{3ex}\\ % \\
WISEA J105633.65$+$574054.5 & 3XMM J105633.4$+$574052 & $0$ & $.48$ & $\pm\,0.09$\rule{0pt}{3ex} & $1$ & $.8020$ & $\pm\,0.0010$\rule{0pt}{3ex} & 2.5 & $($$6$ &  & $\pm\,2$$) \times 10^{-15}$\rule{0pt}{3ex} & $-2$ & $.7$ & $\pm\,0.2$\rule{0pt}{3ex}\\ % \\
WISEA J141821.72$+$522955.2 & CXO J141821.7$+$522955 & $0$ & $.19$ & $\pm\,0.04$\rule{0pt}{3ex} & $1$ & $.5998$ & $\pm\,0.0005$\rule{0pt}{3ex} & 0.4 & $($$4$ &  & $\pm\,2$$) \times 10^{-15}$\rule{0pt}{3ex} & $-2$ & $.7$ & $\pm\,0.3$\rule{0pt}{3ex}\\ % \\
WISEA J160956.04$+$543646.9 & CXO J160956.0$+$543646 & $0$ & $.23$ & $\pm\,0.04$\rule{0pt}{3ex} & $48$ & $.7$ & $\pm\,0.4$\rule{0pt}{3ex} & 0.1 & $($$9$ &  & $\pm\,4$$) \times 10^{-16}$\rule{0pt}{3ex} & $-3$ & $.5$ & $\pm\,0.3$\rule{0pt}{3ex}\\ % \\
WISEA J161112.96$+$541508.3 & 3XMM J161112.8$+$541508 & $0$ & $.25$ & $\pm\,0.04$\rule{0pt}{3ex} & $15$ & $.14$ & $\pm\,0.04$\rule{0pt}{3ex} & 1.0 & $($$2$ & $.5$ & $\pm\,1.0$$) \times 10^{-14}$\rule{0pt}{3ex} & $-1$ & $.9$ & $\pm\,0.3$\rule{0pt}{3ex}\\ % \\
WISEA J221509.36$+$004357.4 & 3XMM J221509.2$+$004356 & $0$ & $.16$ & $\pm\,0.03$\rule{0pt}{3ex} & $1$ & $.1948$ & $\pm\,0.0002$\rule{0pt}{3ex} & 1.8 & $($$1$ & $.6$ & $\pm\,1.0$$) \times 10^{-14}$\rule{0pt}{3ex} & $-2$ & $.6$ & $\pm\,0.5$\rule{0pt}{3ex}\\ % \\
WISEA J221513.23$-$004829.3 & CXO J221513.1$-$004829 & $0$ & $.51$ & $\pm\,0.09$\rule{0pt}{3ex} & $1$ & $.372$ & $\pm\,0.004$\rule{0pt}{3ex} & 0.7 & $($$2$ & $.1$ & $\pm\,0.5$$) \times 10^{-14}$\rule{0pt}{3ex} & $-2$ & $.50$ & $\pm\,0.10$\rule{0pt}{3ex}\\ % \\
WISEA J221722.17$-$002632.9 & 3XMM J221722.1$-$002633 & $0$ & $.09$ & $\pm\,0.02$\rule{0pt}{3ex} & $2$ & $.401$ & $\pm\,0.002$\rule{0pt}{3ex} & 0.8 & $($$1$ & $.7$ & $\pm\,1.0$$) \times 10^{-14}$\rule{0pt}{3ex} & $-2$ & $.2$ & $\pm\,0.4$\rule{0pt}{3ex}\\ %
\enddata
\tablecomments{Column (5) is the separation between the PS1-MDS source position and the
    cataloged X-ray position. Column (6) is the X-ray flux in the 0.2--2~keV
    band.}
\end{deluxetable*}
% end TableBuilder table

%% file: mann.tex
% TableBuilder table
\begin{deluxetable}{ccccccr@{}lr@{}lr@{}lr@{}lr@{}lr@{}lr@{}lr@{}l}
%custom preamble

%hardcoded preamble
\tablecolumns{22}
\tablewidth{0em}
\tablecaption{Coefficients for Estimating Stellar Properties\label{t.mann}}
\tablehead{
\colhead{$y$} & \colhead{Unit} & \colhead{$C_a$} & \colhead{$C_{a,\text{min}}$} & \colhead{$C_{a,\text{max}}$} & \colhead{$C_b$} & \multicolumn{2}{c}{$a_0$} & \multicolumn{2}{c}{$a_1$} & \multicolumn{2}{c}{$a_2$} & \multicolumn{2}{c}{$a_3$} & \multicolumn{2}{c}{$a_4$} & \multicolumn{2}{c}{$b_1$} & \multicolumn{2}{c}{$b_2$} & \multicolumn{2}{c}{Scatter} \\ \\
\multicolumn{1}{c}{(1)} & \multicolumn{1}{c}{(2)} & \multicolumn{1}{c}{(3)} & \multicolumn{1}{c}{(4)} & \multicolumn{1}{c}{(5)} & \multicolumn{1}{c}{(6)} & \multicolumn{2}{c}{(7)} & \multicolumn{2}{c}{(8)} & \multicolumn{2}{c}{(9)} & \multicolumn{2}{c}{(10)} & \multicolumn{2}{c}{(11)} & \multicolumn{2}{c}{(12)} & \multicolumn{2}{c}{(13)} & \multicolumn{2}{c}{(14)}
}
\startdata
\teff & 3500~K & $(g{-}i)_\text{P1}$ & $1.6$ & $3.5$ &  & $2$ & $.309$ & $-1$ & $.3338$ & $0$ & $.5498$ & $-0$ & $.119$ & $0$ & $.01029$ & & & & & $55$ & $.$\\ % \\
 &  & $(r{-}i)_\text{P1}$ & $0.5$ & $2.0$ &  & $1$ & $.483$ & $-0$ & $.9843$ & $0$ & $.8204$ & $-0$ & $.3733$ & $0$ & $.06376$ & & & & & $56$ & $.$\\ % \\
 &  & $(i{-}z)_\text{P1}$ & $0.3$ & $1.0$ &  & $1$ & $.567$ & $-2$ & $.3613$ & $3$ & $.928$ & $-3$ & $.5069$ & $1$ & $.191$ & & & & & $59$ & $.$\\ % \\
 &  & $(r{-}y)_\text{P1}$ & $0.9$ & $3.5$ &  & $1$ & $.508$ & $-0$ & $.5887$ & $0$ & $.271$ & $-0$ & $.0682$ & $0$ & $.006544$ & & & & & $58$ & $.$\\ % \\
 &  & $(g{-}i)_\text{P1}$ & $1.6$ & $3.5$ & $(g{-}r)_\text{P1}$ & $2$ & $.1$ & $-0$ & $.6569$ & $0$ & $.1422$ & $-0$ & $.01241$ & & & $-0$ & $.3722$ & $0$ & $.1702$ & $53$ & $.$\\ % \\
 &  & $(r{-}i)_\text{P1}$ & $0.5$ & $2.0$ & $(g{-}r)_\text{P1}$ & $2$ & $.094$ & $-0$ & $.4136$ & $0$ & $.1308$ & $-0$ & $.02019$ & & & $-1$ & $.153$ & $0$ & $.4175$ & $53$ & $.$\\ % \\
 &  & $(i{-}z)_\text{P1}$ & $0.3$ & $1.0$ & $(g{-}r)_\text{P1}$ & $2$ & $.186$ & $-0$ & $.9242$ & $0$ & $.577$ & $-0$ & $.1625$ & & & $-1$ & $.245$ & $0$ & $.452$ & $54$ & $.$\\ % \\
 &  & $(r{-}y)_\text{P1}$ & $0.9$ & $3.5$ & $(g{-}r)_\text{P1}$ & $2$ & $.061$ & $-0$ & $.2177$ & $0$ & $.03055$ & $-0$ & $.001817$ & & & $-1$ & $.133$ & $0$ & $.4164$ & $53$ & $.$\\ % \\
\hline Mass & \msun & $(g{-}i)_\text{P1}$ & $1.6$ & $3.5$ &  & $0$ & $.05652$ & $1$ & $.319$ & $-0$ & $.7755$ & $0$ & $.1156$ & & & & & & & $0$ & $.17$\tablenotemark{a}\\ % \\
 &  & $(r{-}i)_\text{P1}$ & $0.5$ & $2.0$ &  & $0$ & $.7666$ & $-0$ & $.1287$ & $-0$ & $.3029$ & $0$ & $.1012$ & & & & & & & $0$ & $.19$\tablenotemark{a}\\ % \\
 &  & $(i{-}z)_\text{P1}$ & $0.3$ & $1.0$ &  & $0$ & $.8164$ & $-0$ & $.3494$ & $-1$ & $.3962$ & $1$ & $.041$ & & & & & & & $0$ & $.19$\tablenotemark{a}\\ % \\
 &  & $(r{-}y)_\text{P1}$ & $0.9$ & $3.5$ &  & $0$ & $.7807$ & $-0$ & $.075$ & $-0$ & $.1147$ & $0$ & $.02327$ & & & & & & & $0$ & $.19$\tablenotemark{a}\\ % \\
 &  & $(g{-}i)_\text{P1}$ & $1.6$ & $3.5$ & $(g{-}r)_\text{P1}$ & $2$ & $.58$ & $0$ & $.4904$ & $-0$ & $.4053$ & $0$ & $.06325$ & & & $-3$ & $.001$ & $1$ & $.15$ & $0$ & $.17$\tablenotemark{a}\\ % \\
 &  & $(r{-}i)_\text{P1}$ & $0.5$ & $2.0$ & $(g{-}r)_\text{P1}$ & $4$ & $.661$ & $-0$ & $.2381$ & $-0$ & $.1947$ & $0$ & $.07523$ & & & $-5$ & $.993$ & $2$ & $.298$ & $0$ & $.17$\tablenotemark{a}\\ % \\
 &  & $(i{-}z)_\text{P1}$ & $0.3$ & $1.0$ & $(g{-}r)_\text{P1}$ & $5$ & $.134$ & $-1$ & $.306$ & $0$ & $.2464$ & $0$ & $.209$ & & & $-6$ & $.393$ & $2$ & $.437$ & $0$ & $.18$\tablenotemark{a}\\ % \\
 &  & $(r{-}y)_\text{P1}$ & $0.9$ & $3.5$ & $(g{-}r)_\text{P1}$ & $4$ & $.448$ & $-0$ & $.4573$ & $0$ & $.06287$ & $-0$ & $.001415$ & & & $-5$ & $.274$ & $2$ & $.014$ & $0$ & $.18$\tablenotemark{a}\\ % \\
\hline $\text{BC}_{z,\text{P1}}$ & mag & $(g{-}i)_\text{P1}$ & $1.6$ & $3.5$ &  & $-0$ & $.28032$ & $0$ & $.74917$ & $-0$ & $.21769$ & $0$ & $.011735$ & & & & & & & $0$ & $.041$\\ % \\
 &  & $(r{-}i)_\text{P1}$ & $0.5$ & $2.0$ &  & $0$ & $.27469$ & $0$ & $.39134$ & $-0$ & $.23959$ & $0$ & $.015105$ & & & & & & & $0$ & $.04$\\ % \\
 &  & $(i{-}z)_\text{P1}$ & $0.3$ & $1.0$ &  & $0$ & $.20833$ & $1$ & $.1028$ & $-1$ & $.4065$ & $0$ & $.25586$ & & & & & & & $0$ & $.037$\\ % \\
 &  & $(r{-}y)_\text{P1}$ & $0.9$ & $3.5$ &  & $0$ & $.25276$ & $0$ & $.25951$ & $-0$ & $.094812$ & $0$ & $.0004336$ & & & & & & & $0$ & $.038$\\ % \\
$\text{BC}_{r,\text{P1}}$ & mag & $(r{-}y)_\text{P1}$ & $0.9$ & $3.5$ &  & $0$ & $.2528$ & $-0$ & $.5874$ & $-0$ & $.1209$ & $0$ & $.009721$ & & & & & & & $0$ & $.03$\\ % \\
$\text{BC}_{i,\text{P1}}$ & mag & $(r{-}i)_\text{P1}$ & $0.5$ & $2.0$ &  & $0$ & $.2728$ & $-0$ & $.1441$ & $-0$ & $.1914$ & $-0$ & $.001051$ & & & & & & & $0$ & $.047$\\ % \\
$\text{BC}_{y,\text{P1}}$ & mag & $(r{-}y)_\text{P1}$ & $0.9$ & $3.5$ &  & $0$ & $.2707$ & $0$ & $.3889$ & $-0$ & $.11$ & $0$ & $.008266$ & & & & & & & $0$ & $.03$\\ % \\
\hline $\text{M}_{z,\text{P1}}$ & mag & $(g{-}i)_\text{P1}$ & $1.6$ & $3.5$ &  & $8$ & $.116$ & $-4$ & $.5982$ & $3$ & $.025$ & $-0$ & $.4002$ & & & & & & & $0$ & $.37$\\ % \\
 &  & $(r{-}i)_\text{P1}$ & $0.5$ & $2.0$ &  & $6$ & $.211$ & $0$ & $.7751$ & $1$ & $.81$ & $-0$ & $.4355$ & & & & & & & $0$ & $.41$\\ % \\
 &  & $(i{-}z)_\text{P1}$ & $0.3$ & $1.0$ &  & $5$ & $.969$ & $1$ & $.636$ & $8$ & $.947$ & $-4$ & $.7744$ & & & & & & & $0$ & $.42$\\ % \\
 &  & $(r{-}y)_\text{P1}$ & $0.9$ & $3.5$ &  & $6$ & $.046$ & $0$ & $.5451$ & $0$ & $.6392$ & $-0$ & $.0961$ & & & & & & & $0$ & $.41$\\ % \\
 &  & $(g{-}i)_\text{P1}$ & $1.6$ & $3.5$ & $(g{-}r)_\text{P1}$ & $4$ & $.38$ & $-23$ & $.51$ & $14$ & $.2$ & $-3$ & $.309$ & $0$ & $.2817$ & $24$ & $.89$ & $-9$ & $.885$ & $0$ & $.36$\\ % \\
 &  & $(r{-}i)_\text{P1}$ & $0.5$ & $2.0$ & $(g{-}r)_\text{P1}$ & $-14$ & $.26$ & $-3$ & $.695$ & $6$ & $.901$ & $-3$ & $.125$ & $0$ & $.5108$ & $32$ & $.05$ & $-11$ & $.38$ & $0$ & $.37$\\ % \\
 &  & $(i{-}z)_\text{P1}$ & $0.3$ & $1.0$ & $(g{-}r)_\text{P1}$ & $-15$ & $.77$ & $-9$ & $.422$ & $33$ & $.9$ & $-30$ & $.85$ & $9$ & $.942$ & $34$ & $.79$ & $-12$ & $.54$ & $0$ & $.36$\\ % \\
 &  & $(r{-}y)_\text{P1}$ & $0.9$ & $3.5$ & $(g{-}r)_\text{P1}$ & $-16$ & $.25$ & $-1$ & $.54$ & $1$ & $.906$ & $-0$ & $.4622$ & $0$ & $.03956$ & $35$ & $.38$ & $-13$ & $.04$ & $0$ & $.36$\\ %
\enddata
\tablecomments{Each row provides values for evaluating a polynomial having the form of
\autoref{e.mann_polynomial}. Columns (4) and (5) give the range of values for
which each relation should be used. Column (14) gives the rms scatter of the
polynomial fit to the \cite{mann2015} sample.}
\tablenotetext{a}{Scatter is fractional rather than absolute.}
\end{deluxetable}
% end TableBuilder table